\documentclass[longauth]{aa}  
\usepackage{graphicx}
\usepackage{txfonts}
\usepackage{natbib}
\usepackage{longtable}
\usepackage{supertabular}
\bibpunct{(}{)}{;}{a}{}{,}
\newcommand{\tn}[1]{\textnormal{#1}}

\begin{document}
   \title{Highly luminous supernovae associated with gamma-ray bursts I.: GRB 111209A/SN 2011kl in the context of stripped-envelope and superluminous supernovae\thanks{Partially based on observations obtained under programme 088.A-0051(C), PI: J. P. U. Fynbo}}

   \author{
D. A. Kann		\inst{	1,2,3,4	} \and
P. Schady		\inst{	2	} \and
F. Olivares E.		\inst{	5	} \and
S. Klose		\inst{	1	} \and
A. Rossi		\inst{	6,1	} \and
D. A. Perley		\inst{	7,8,9	} \and
T. Kr\"uhler		\inst{	2,8,10	} \and
J. Greiner		\inst{	2,3	} \and
A. Nicuesa Guelbenzu		\inst{	1	} \and
J. Elliott		\inst{	2,11	} \and
F. Knust		\inst{	2	} \and
R. Filgas		\inst{	12	} \and
E. Pian		\inst{	5,13	} \and
P. Mazzali		\inst{	14,15	} \and
J. P. U. Fynbo		\inst{	8	} \and
G. Leloudas		\inst{	16,8	} \and
P. M. J. Afonso		\inst{	17	} \and
C. Delvaux		\inst{	2	} \and
J. F. Graham		\inst{	2	} \and
A. Rau		\inst{	2	} \and
S. Schmidl		\inst{	1	} \and
S. Schulze		\inst{	18,19,20	} \and
M. Tanga		\inst{	2	} \and
A. C. Updike		\inst{	21	} \and
K. Varela		\inst{	2	}
 }

   \offprints{D. A. Kann, \email{kann@iaa.es}}

\institute{
Th\"uringer Landessternwarte Tautenburg, Sternwarte 5, 07778 Tautenburg, Germany 
\and
Max-Planck-Institut f\"ur extraterrestrische Physik, Giessenbachstra\ss e 1, 85748 Garching, Germany 
\and
Universe Cluster, Technische Universit\"at M\"unchen, Boltzmannstra\ss e 2, 85748 Garching, Germany 
\and
Instituto de Astrof\'isica de Andaluc\'ia (IAA-CSIC), Glorieta de la Astronom\'ia s/n, 18008 Granada, Spain 
\and
Departamento de Ciencias Fisicas, Universidad Andres Bello, Avda. Republica 252, Santiago, Chile 
\and
INAF-IASF Bologna, Area della Ricerca CNR, via Gobetti 101, I--40129 Bologna, Italy 
\and
Cahill Center for Astrophysics, California Institute of Technology, Pasadena, CA 91125, USA 
\and
Dark Cosmology Centre, Niels Bohr Institute, University of Copenhagen, Juliane Maries Vej 30, 2100 Copenhagen, Denmark 
\and
Astrophysics Research Institute, Liverpool John Moores University, IC2, Liverpool Science Park, 146 Brownlow Hill, Liverpool L3 5RF, UK 
\and
ESO, Alonso de Cordova 3107, Vitacura, Santiago de Chile, Chile. 
\and
Harvard-Smithsonian Center for Astrophysics, 60 Garden St., Cambridge, MA 02138, USA 
\and
Institute of Experimental and Applied Physics, Czech Technical University in Prague, Horsk\'a 3a/22, 12800 Prague, Czech Republic 
\and
Scuola Normale Superiore, Piazza dei Cavalieri 7, 56126 Pisa, Italy 
\and
Astrophysics Research Institute, Liverpool John Moores University, 146 Brownlow Hill, Liverpool L3 5RF, UK 
\and
Max-Planck Institut f\"ur Astrophysik, Karl-Schwarzschild-Str. 1, 85748 Garching, Germany 
\and
Department of Particle Physics \& Astrophysics, Weizmann Institute of Science, Rehovot 76100, Israel. 
\and
American River College, Physics and Astronomy Dpt., 4700 College Oak Drive, Sacramento, CA 95841, USA 
\and
Instituto de Astrof\'isica, Facultad de F\'isica, Pontificia Universidad Cat\'olica de Chile, Av. Vicu\~na Mackenna 4860, 306, Santiago 22, Chile. 
\and
Millennium Institute of Astrophysics, Av. Vicu\~na Mackenna 4860, 7820436 Macul, Santiago, Chile. 
\and
Department of Particle Physics and Astrophysics, Faculty of Physics, Weizmann Institute of Science, Rehovot 76100, Israel 
\and
Department of Chemistry and Physics, Roger Williams University, One Old Ferry Road, Bristol, RI 02809, USA 
}

\date{Received 22 June 2016 / Accepted }


\begin{abstract}
{GRB 111209A, {one of the longest Gamma-Ray Bursts (GRBs) ever observed}, is linked to SN 2011kl, the most luminous GRB-Supernova (SN) detected so far, which shows evidence for being powered by a magnetar central engine.}
{We place SN 2011kl into the context of large samples of SNe, addressing in more detail the question of whether it could be radioactively powered, and whether it represents an extreme version of a GRB-SN or an underluminous Superluminous SN (SLSN).}
{We model SN 2011kl using SN 1998bw as a template and derive a bolometric light curve including near-infrared data. We compare the properties of SN 2011kl to literature results on stripped-envelope and superluminous supernovae.}
{Comparison in the $k,s$ context, i.e., comparing it to SN 1998bw templates in terms of luminosity and light-curve stretch, clearly shows SN 2011kl is the most luminous GRB-SN to date, and it is spectrally very dissimilar to other events, being significantly bluer/hotter. Although SN 2011kl does not reach the {classical luminosity threshold of} SLSNe and evolves faster than any of them, it resembles SLSNe more than the classical GRB-associated broad-lined Type Ic SNe in several aspects.}
{GRB 111209A was a very energetic event, both at early (prompt emission) and at very late (SN) times. We have shown in a further publication that with the exception of the extreme duration, the GRB and afterglow parameters are in agreement with the known distributions for these parameters. SN 2011kl, on the other hand, is exceptional both in luminosity and spectral characteristics, indicating that GRB 111209A was likely not powered by a standard-model collapsar central engine, further supporting our earlier conclusions. Instead, it reveals the possibility of a direct link between GRBs and SLSNe.}
\end{abstract}


\authorrunning{D. A. Kann et al.}
\titlerunning{GRB 111209A/SN 2011kl in context}
\maketitle
%

\section{Introduction}
\label{SectInt}
Gamma-Ray Bursts (GRBs) are the most luminous explosions in the Universe (see, e.g., \citealt{Gehrels2009ARAA} for a recent review). Their afterglow emission can be extremely luminous during and right after the GRB \citep{Kann2007AJ,Racusin2008Nature,Bloom2009ApJ}. There are at least two classes of GRBs \citep{Mazets1981ApSS,Kouveliotou1993ApJ}, and the class generally known as long GRBs (or Type II GRBs in a more physically motivated classification scheme which is independent of duration, \citealt{Gehrels2006Nature,Zhang2007ApJ,Zhang2009ApJ,Kann2010ApJ,Kann2011ApJ}) has been shown to be conclusively linked to the supernovae (SNe) explosions of very massive stars (e.g., \citealt{Galama1998Nature}, \citealt{Hjorth2003Nature}, \citealt{Stanek2003ApJ}, see, e.g.,
\citealt{Cano2016JAA} for a review, and \citealt{Barnes2018ApJ} for numerical modelling). These stars are likely Wolf-Rayet stars, which are thought to be linked to Type Ic SNe, the explosions of highly stripped massive stellar cores which have {either ejected (via binary interaction up to common-envelope phases, e.g., \citealt{Fryer2005ApJ,Sana2012Science}) or burned (via chemically homogeneous evolution, e.g., \citealt{Yoon2005AA})} their H and He envelopes (see, e.g., \citealt{Smartt2009ARAA} for a review){. The advent of untargeted automatic sky surveys has led to the discovery of large numbers of these so-called stripped-envelope SNe \citep[e.g.,][]{Taddia2018AA,Taddia2019AA2,Stritzinger2018AA1,Stritzinger2018AA2}
, allowing statistically significant studies of the distributions of luminosity or rise/decay times, spectral characteristics (e.g., the question of He in Type Ic spectra), expansion speeds, produced $^{56}$Ni masses, ejecta kinetic energies, and more.}
Note that \cite{Sobacchi2017MNRAS2} posit that the existence of a He layer prevents jet breakout, therefore GRBs are not associated with Type Ib SNe
. Such highly stripped SNe generally exhibit very high expansion velocities of $\approx0.1$ c {(e.g., \citealt{Bufano2012ApJ,Schulze2014AA,Izzo2019Nature}),}
leading to the term ``broad-lined Type Ic SNe'' (henceforth Type Ic-BL SNe). Recently, \cite{Prentice2017MNRAS} presented a physically motivated classification of SE-SNe, they find all GRB-SNe are ``Type Ic-3'' (conversely, not all Type Ic SNe with high expansion speeds are associated with GRBs, and almost all of these have $N>3$ in the classification scheme of \citealt{Prentice2017MNRAS}). In rare cases, such Type Ic-BL SNe show evidence for relativistic ejecta but without an associated GRB, e.g., {SN 2009bb \citep{Soderberg2010Nat,Pignata2011ApJ}, SN 2012ap \citep{Margutti2014ApJ,Milisavljevic2015ApJ},} and iPTF17cw \citep{Corsi2017ApJ}, the latter possibly being associated with a GRB.

The GRB 111209A, discovered by the Neil Gehrels \emph{Swift} Observatory satellite \citep{Gehrels2004ApJ}, is a truly remarkable event, at 25 ks the {second-longest (after GRB 170714A, \citealt{Dai2017GCN,Kann2017GCN})} GRB ever discovered \citep{Golenetskii2011GCN,Gendre2013ApJ}, and one of the very rare ultra-long GRBs (ULGRBs, \citealt{Levan2014ApJ}, henceforth L14; see also \citealt{Levan2015Rome}). Using the GROND instrument \citep{Greiner2008PASP}, we discovered that GRB 111209A was accompanied by a very luminous SN \citep[][henceforth G15]{Greiner2015Nat}, dubbed SN 2011kl, which is the most luminous GRB-SN discovered so far, and spectrally dissimilar to any known GRB-SN, being much bluer and hotter, and exhibiting a spectrum much more in accordance with those of superluminous supernovae (SLSNe; G15, \citealt{Mazzali2016MNRAS}, \citealt{Liu2017ApJ}). The afterglow of GRB 111209A shows a complex evolution but is generally unremarkable within the context of GRB afterglows (\citealt{Stratta2013ApJ}, \citealt{Kann2017AA_GRB111209A}, henceforth K18B).

In this paper, we build upon our earlier results (G15,K18B). We derive a bolometric light curve of SN 2011kl incorporating an IR correction, which had not been undertaken by G15. Using both our own extensive GRB-SN analysis (Kann et al., in prep.) as well as large samples derived from the literature, we place SN 2011kl in context and study whether this is an extreme GRB-SN, or is more similar to SLSNe, which would establish a direct connection between the most luminous SNe and the most luminous high-energy transients. 

The paper is organized as follows: In Sect. \ref{SectFit} we present our fit to the late-time data of the afterglow of GRB 111209A (further to the analysis and results of G15 and K18B) and the derivation of the parameters of SN 2011kl as well as the bolometric light curve. In Sect. \ref{SectRes} we present results on the blackbody fits of the SN emission as well as the absolute magnitudes of the SN. In Sect. \ref{SectDisc}, we place SN 2011kl into the context of GRB-SNe, other SE-SNe and SLSNe, and we discuss the nature of GRB 111209A/SN 2011kl in the light of our combined results. We reach our conclusions in Sect. \ref{SectConc}.

We will follow the convention $F_\nu\propto t^{-\alpha}\nu^{-\beta}$ to describe the temporal and spectral evolution of the afterglow. We use WMAP $\Lambda$CDM concordance cosmology \citep{Spergel2003ApJS} with $H_0=71$km s$^{-1}$ Mpc$^{-1}$, $\Omega_{\rm M}=0.27$, and $\Omega_{\Lambda}=0.73$. Uncertainties are given at 68\% ($1\sigma$) confidence level for one parameter of interest unless stated otherwise, whereas upper limits are given at the $3\sigma$ confidence level.


\section{Fitting SN 2011kl}
\label{SectFit}

Using a combination of our unique GROND data set, as well as crucial data from the literature, G15 for the first time found evidence for both a jet break as well as a late-time supernova component for this GRB. This jet break is actually hidden by the rising SN in all filters except $U$. (We also note we have a data gap between 13 and 22 days that is larger on the logarithmic scale than the time between other data points due to a period of bad weather, we may therefore have missed a steeper decay component during this time.) The consensus in the literature prior to the publication of our results in G15 had been that this GRB exhibited the \textit{lack of an associated supernova}, which drove much of the interpretation and modelling of the data that had been published at that time (\citealt{Gendre2013ApJ}, L14, \citealt{Stratta2013ApJ,Kashiyama2013ApJ,Nakauchi2013ApJ}, see Sect. \ref{SectDiscNature}). Our detection therefore was equal to a paradigm shift. We describe the fitting procedure in detail in the following.

\subsection{Fitting the late afterglow and the supernova}
\label{secSN}

\begin{figure*}
  \centering
 \includegraphics[width=\textwidth]{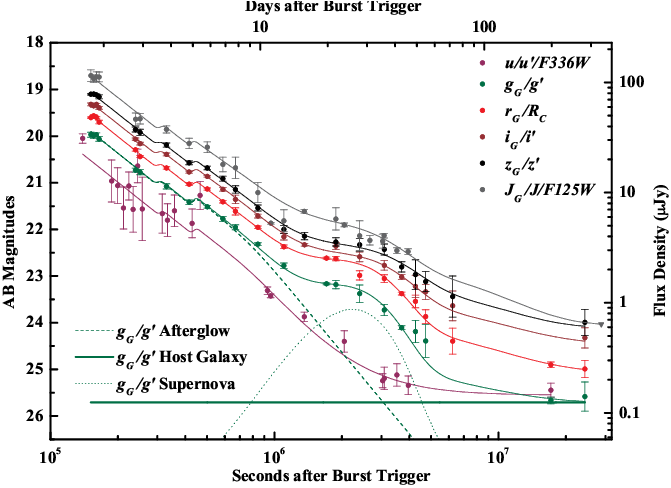}
   \caption{The late light curve of the optical/NIR afterglow of GRB 111209A in $Ug^\prime r^\prime i^\prime z^\prime J$ fit ($U$ being UVOT $u$, SDSS $u^\prime$ and \emph{HST} $F336W$) with the combination of a broken power-law afterglow component (including two steps at earlier times), a SN component based on the SN 1998bw template, and a late-time host-galaxy value. The $U$ light curve does not incorporate a SN and has been used to determine the (jet) break time and post-break decay slope. The full line is the fit to each colour. For the $g^\prime$ band, we additionally show the decomposition into afterglow, host galaxy and SN component. The fit shown is identical to the one given in G15, here we show the best-fit curves for all filters.}
              \label{SNfig}
    \end{figure*}

The GROND/UVOT data of the afterglow of GRB 111209A were given in K18B, where we also present the data analysis and the complete analysis of the early afterglow. From $\approx6$ days to $\approx10$ days, and possibly even to $\approx13$ days, the afterglow decays monotonically and achromatically, before showing a strong departure from this behaviour at $>20$ days. This ``late red bump'' is a characteristic sign of a late-time supernova contribution to the optical transient following a GRB \citep[e.g.,][]{Price2003ApJ}. In the $white$ data as well as the $F336W/u^\prime$ data from L14, the afterglow is seen to decay more rapidly.

Such bumps were already found in the very first afterglows, such as for GRB 970228 \citep{Galama2000ApJ} and GRB 980326 \citep{Bloom1999Nature}. They were studied systematically by \cite{Zeh2004ApJ}, who came to the conclusion that all afterglows of GRBs at $z\leq0.7$ (as well as some in the range $0.7\leq z\leq1.1$) showed evidence of a late-time SN contribution. Depending on the evolution of the afterglow, these bumps are more or less easy to detect. In case of an afterglow with an early break and a steep decay, as well as a faint (or well-subtracted) host contribution, the bumps are very prominent (e.g., GRB 011121: \citealt{Garnavich2003ApJ, Greiner2003ApJ}; GRB 020405: \citealt{Price2003ApJ, Masetti2003AA}; GRB 041006: \citealt{Stanek2005ApJ}). If the afterglow decay is less steep, this results into a transition to a plateau phase which can be mistaken for the constant host galaxy contribution, but will show a further drop after $\approx50...100$ days (e.g., GRB 050525A: \citealt{DellaValle2006ApJ}; GRB 090618: \citealt{Cano2011MNRAS}; GRB 091127: \citealt{Cobb2010ApJ,Filgas2011AA,Vergani2011AA}). Finally, in case of a distant GRB with a host galaxy of similar magnitude (e.g., GRB 021211: \citealt{DellaValle2003AA}), or a bright, slowly decaying afterglow (e.g, GRB 030329: \citealt{Kann2006ApJ}; XRF 050824: \citealt{Sollerman2007AA}; GRB 060729: \citealt{Cano2011MNRAS}), the contribution of the SN can be hard to decipher.

We perform the supernova fitting together with that of the late afterglow using the method of \cite{Zeh2004ApJ} and giving the results in the $k,s$ formalism (see below). Those authors used the light curves of SN 1998bw as given by \cite{Galama1998Nature} as a template, and derived an analytical equation which is able to fit the data well \citep{Klose2018AA}. We create SN 1998bw template light curves in the GROND bandpasses (as well as $J$ from L14) and at the redshift of GRB 111209A, i.e., the resulting light curves show what SN 1998bw would look like at $z=0.677$ if observed with GROND/HAWK-I $J$. These templates are fitted with the above-mentioned analytical equation, which can then be used to perform non-linear curve fitting using $\chi^2$ minimization on the data of GRB 111209A. The SN 1998bw template is unique in each bandpass and described by a set of seven parameters, which are different depending on the bandpass and (in case of other GRBs) the redshift \citep{Zeh2004ApJ}. The entire fit uses a joint afterglow model with a single decay index (there is no more significant evidence for chromatic behaviour as is seen in the rebrightening, see K18B), but five differently-scaled templates for the SN-fitting. These templates are additionally modified by the $k$ and $s$ values. The luminosity factor $k$ describes how luminous the GRB-SN is relative to SN 1998bw itself, i.e., it scales the entire light curve up and down. The stretch factor $s$ is simply a multiplicative factor applied to the time axis; while we assume the general behaviour of the light curve remains identical (and the SN evolution initializes at the time of the GRB itself), i.e., $s$ is able to make the evolution faster or slower than SN 1998bw, which by definition has $k=1,s=1$. So far, while GRB-SNe have been found that evolve considerably faster than SN 1998bw \citep{Ferrero2006AA, Sollerman2007AA, Olivares2012AA} and also considerably slower \citep{Stanek2005ApJ}, none have showed significant deviations from the template shape, validating the formalism. This is definitely \emph{not} true for Type Ic SNe in general, even BL ones not associated with GRBs, see the appendix of \cite{Ferrero2006AA}.

The afterglow and SN phase are excellently covered by GROND in $g^\prime r^\prime i^\prime z^\prime $, but the SN is not detected in $J_{\rm GROND}$. Therefore, we add VLT/$J$ and \emph{HST}/$F125W$ data from L14, except for the first \emph{HST} $F125W$ point, which is anomalously faint (see Fig. \ref{SNfig} at 960 ks, on the $i^\prime$-band fit curve). L14 also have two epochs of $g^\prime R_C i^\prime z^\prime $ data during the SN phase, which agree excellently with our data, as L14 used our GROND calibration stars. Finally, the most important contribution from L14 is late-time $u^\prime$ and \emph{HST} $F336W$ data. These authors find a decay slope of $\alpha=1.38$ from combined UVOT $u$ and their $u^\prime F336W$ data. This value is in decent agreement with our values determined from our joint afterglow fit, but L14 fit a single power-law to data from $\approx1-50$ days. Their error while performing this fit lies in ignoring a host-galaxy contribution; the data at $\approx30-50$ days is already on the level of the host-galaxy (which is actually noted in their work). Taking this into account, we see clear evidence for a steeper afterglow decay at late times. As discussed in more detail in Sect. \ref{SectDiscSN}, we do not expect a SN contribution at this redshift in this wavelength range (which we call $U$ for convenience from now on, it is roughly analogue to the rest-frame UVOT $uvw2$ band, at 1928 {\AA}), something that is also fully validated by the SN spectrum (G15), therefore we use these data as a template for the pure afterglow contribution. To obtain a good fit, we combine UVOT $u$, Gemini $u^\prime$ and \emph{HST} $F336W$ data into one light curve. While these filters differ slightly, the data are dominated by statistical errors, so we do not expect small filter mismatches to affect the result as long as all values are in AB magnitudes. To obtain the best possible value for the pre-break afterglow slope, we use data starting at the second GROND epoch (the beginning of the post-rebrightening decay), taking the two ``steps'' into account and fixing the decay slope to be identical in all three segments (this is motivated by our interpretation of these features, see K18B, and we find no evidence for a strong discrepancy compared to the null hypothesis of a single decay slope). Finally, we add a filter-dependent host-galaxy component at very late times, both from our data (see K18B for further discussion of the host galaxy) as well as from L14 (note we use their \emph{HST} $F336W$ value for the $U$ band). The complete fit therefore consists of the following components and parameters:

\begin{itemize}
\item An afterglow component, described by a broken power-law, with pre-break decay slope $\alpha_{late,1}$, post-break decay slope $\alpha_{late,2}$, break time $t_b$ and break smoothness $n$ as shared parameters for all filters, while the normalization $m_k$ (magnitude at $t_b$ assuming $n=\infty$) is an individual parameter for each filter.
\item A supernova component in all bands except $U$, with luminosity factor $k$ and stretch factor $s$ being individual parameters for each filter.
\item A host galaxy component, individual for each filter, for $J$, we use the upper limit from L14, which we have independently confirmed, as a maximum possible host value.
\end{itemize}

Our fit result is shown in Fig. \ref{SNfig}. Due to some scatter in the data (e.g., the early UVOT $u$ data), the fit is not quite satisfactory ($\chi^2=143.5$ for 109 degrees of freedom). For the afterglow, we find $\alpha_{late,1}=1.55\pm0.01$, $\alpha_{late,2}=2.33\pm0.16$, $t_b=9.12\pm0.47$ days, and $n=10$ has been fixed. The parameters for the SN are given in Table \ref{tab_fit4}. This fit was also used in G15.

The only other study which has presented late-time observations is L14. While they have data during the SN phase which is, as pointed out further above, in excellent agreement with our own, they are lacking high S/N observations of the afterglow between $\approx1$ day (where they have further Gemini observations, a single epoch in several colours, during the steep rise) and the time when the SN dominates. This leads to fits with different decay slopes (in $U$ and $J$, see their Fig. 6), an inability to clearly separate the afterglow and SN components, and only weak constraints on a jet break. L14 do present several tantalizing hints toward the existence of the SN. First, they find clear reddening at late times in comparison to their earlier data. Secondly, next to early VLT/X-shooter and Gemini/GMOS afterglow spectroscopy (which determined the redshift), they also obtained two ground-based (again, Gemini/GMOS and VLT/X-shooter, at 11 and 20 days post-burst, respectively) and two space-based (\emph{HST} WFC3 grism, at 11 and 35 days) spectra, at the beginning and during the SN. Curiously, while confirming the reddening seen in the photometry, these spectra do \textit{not} show clear undulations which would be expected from a typical Type Ic-BL SN associated with a GRB (L14, G15). On the other hand, while redder in comparison to the early afterglow, these spectra were much bluer in the rest-frame UV than expected for a typical GRB-SN (note there is another \emph{possible} SN which shows a similar rather blue and flat SED, namely the late-time bump associated with XRF 030723, though not even a spectroscopic redshift is known for this event, see \cite{Fynbo2004ApJ,Tominaga2004ApJ} and \cite{Huang2004ApJ} for more discussion). We confirm these general results with our own reduction of the X-shooter spectra (\citealt{Kruhler2015AA}, G15). This leads us to our supernova results.

\subsection{The highly luminous SN 2011kl}

While the $s$ values are roughly similar, pointing to an evolution which is somewhat slower than that of SN 1998bw (except for $J$; here, the lack of data during the SN decay may skew the result), the $k$ values diverge quite strongly in colour (independent of its actual luminosity, a SN spectrally identical to SN 1998bw would show identical $k$ values independent of the band), in an almost monotonic fashion, the bluer, the larger ($i^\prime z^\prime $ being identical within errors, $r^\prime $ significantly larger, and $g^\prime $ even larger). Only the $J$ band, once again, deviates from this pattern. Incidentally, L14 point out that if their late-time $J$ data were to be associated with a supernova, it would be superluminous. We also confirm the existence of a flux excess even beyond the extrapolation of the SN spectrum in $JH$ from our reduction of the X-shooter spectrum at 19.8 days, see the Extended Data in G15.

This is in full agreement with our second result: the SN is very luminous in general, especially after we also correct for the (small) line-of-sight extinction (K18B). In $i^\prime z^\prime$, SN 2011kl is $\approx1.8\,\times\,$SN 1998bw in luminosity (0.6 mag brighter), in $r^\prime$ it is $3.1\,\times$ more luminous (1.2 mag), while it is $5\,\times$ as luminous in $g^\prime$ (1.75 mag brighter), an unprecedented result. We caution that the observed $g^\prime$ band corresponds to the ultraviolet (2735 {\AA}, roughly the UVOT \emph{uvw1} band) in the rest-frame, and there are no data for SN 1998bw in this bandpass, therefore the light curve was derived by extrapolation. The observed $r^\prime$ band (rest-frame wavelength 3700 {\AA}), on the other hand, can be directly compared to the SN 1998bw $U$-band data, implying that our extrapolation to get an observer-frame $g^\prime$ light curve must be reasonably robust. These values show that SN 2011kl is not just spectrally significantly dissimilar to SN 1998bw, being much more ultraviolet-luminous, but it is also the most luminous GRB-SN detected so far (G15, Sect. \ref{SectDiscSN}).

A conservative estimate of the minimum luminosity of the SN can be gained by fitting the afterglow with an unbroken power-law, increasing the afterglow contribution to the total optical transient at the time of the SN. This yields a significantly worse fit, it is $\chi^2=183.7$ for 111 degrees of freedom for the simple power-law fit vs. $\chi^2=143.5$ for 109 degrees of freedom for the broken power-law fit, yielding $\Delta\chi^2=40.2$ for two more degrees of freedom. We find a decay slope identical to $\alpha_1$ from the broken power-law fit, and (once again assuming no SN contribution in the $U$ band): $k_{g^\prime}=2.51\pm0.24$, $k_{r^\prime}=2.00\pm0.14$, $k_{i^\prime}=1.02\pm0.16$, $k_{z^\prime}=1.05\pm0.22$, and $k_{J}=2.30\pm0.22$ (and $s$ values larger than those found from the broken power-law fit in the range of 10\%--20\%). These values are significantly less luminous (and unremarkable in $i^\prime z^\prime$), just 50\% -- 64\% of those we find using a broken power-law fit. As stated, though, this fit is significantly worse and can therefore be ruled out in comparison to the broken power-law fit. Such an unbroken power-law would also not be expected (but see \citealt{Perley2014ApJ}, \citealt{DePasquale2016MNRAS2}), and the wide opening angle implied by an extremely late break would increase the energetics of the GRB to a level not accommodated by the magnetar model (but see \citealt{Metzger2015MNRAS}), which is strongly supported by the spectral characteristics (G15). (Also see \citealt{Gompertz2017ApJ} for a similar discussion, and K18B for a discussion of their results in the light of our full data set.)

We initially used the $U$ light curve as a pure afterglow component under the assumption that there would be no contribution from the SN, and now we have found that the UV-damping usually seen for GRB-SNe does not apply to this one, at least down to approximately the (observer-frame) $g^\prime$ band. Therefore, it is possible the SN also contributes somewhat to the $U$-band light curve (in Fig. \ref{SNfig}, the data at 16 and 27 days indeed lie marginally above the fit, though this result is not statistically significant). If this is the case, though, our result is only strengthened. Additional SN light in $U$ implies that the intrinsic post-break decay slope must be even steeper, therefore the afterglow contribution during the SN epoch will be even smaller, and the SN will be even more luminous. Our $k$ results should thus be treated as robust lower limits to the SN luminosity, though we do not expect it to be significantly more luminous than what we have already found.

\begin{table}[t]
\caption{Results of the Supernova Fit.}
\label{tab_fit4}
\centering                        
\begin{tabular}{l c c c c}       
\hline\hline   
Filter	& $\lambda_{rest}$ (\AA) &	$m_k$			&	$k$			&	$s$			\\\hline   
$U$	& 2138 & $	22.955	\pm	0.112	$ & $	\cdots$ & $\cdots$ \\
$g^\prime$	& 2735 & $	22.286	\pm	0.099	$ & $	4.97	\pm	0.49	$ & $	1.26	\pm	0.02	$ \\
$r^\prime$	& 3709 & $	21.970	\pm	0.097	$ & $	3.14	\pm	0.27	$ & $	1.13	\pm	0.02	$ \\
$i^\prime$	& 4556 & $	21.713	\pm	0.096	$ & $	1.81	\pm	0.22	$ & $	1.12	\pm	0.05	$ \\
$z^\prime$	& 5360 & $	21.544	\pm	0.098	$ & $	1.71	\pm	0.25	$ & $	1.05	\pm	0.08	$ \\
$J$	& 7394 & $	21.146	\pm	0.102	$ & $	3.59	\pm	0.33	$ & $	0.79	\pm	0.03 $ \\
\hline \hline
\end{tabular}
\tablefoot{These results are based on a joint fit of all bands showing a SN component as well as the $U$ band from which we derive the pure afterglow parameters. See text for more details.}\\
\end{table}

\onltab{2}{
\longtab{2}{
\begin{longtable}{rcrcccccc}
\caption{\label{SNtabrest} The values from Table 1 in G15, given in the rest-frame.}\\
\hline
\hline                
$\Delta$t$_{obs}$ (s)	&	$\Delta$t$_{obs}$ (d)	&	$\Delta$t$_{rest}$ (s)	&	$\Delta$t$_{rest}$ (d)	&	$M_{2735}$ mag						&	$M_{3709}$ mag					&	$M_{4556}$ mag					&	$M_{5360}$ mag		&	$M_{7394}$ mag					\\ \hline \vspace{1mm}
843664	&	9.7646	&	503073	&	5.8226	&	$	-18.69	^{+	0.26	}_{-	0.21	}	$	&	$	-19.13	^{+	0.23	}_{-	0.19	}	$	&	$	-19.02	^{+	0.55	}_{-	0.38	}	$	&	$	-19.08	^{+	1.13	}_{-	0.57	}	$	&	$	\cdots	$	\\	\vspace{1mm}
1101930	&	12.7538	&	657076	&	7.6050	&	$	-18.88	^{+	0.29	}_{-	0.24	}	$	&	$	-19.39	^{+	0.16	}_{-	0.14	}	$	&	$	-19.25	^{+	0.44	}_{-	0.33	}	$	&	$	-19.22	^{+	0.75	}_{-	0.48	}	$	&	$	\cdots	$	\\	\vspace{1mm}
1358649	&	15.7251	&	810157	&	9.3768	&	$	\cdots	$	&	$	\cdots	$	&	$	\cdots	$	&	$	\cdots	$	&	$	-20.67	\pm	0.09	$	\\	\vspace{1mm}																		
1360463	&	15.7461	&	811238	&	9.3893	&	$	\cdots	$	&	$	\cdots	$	&	$	-19.77	^{+	0.12	}_{-	0.11	}	$	&	$	\cdots	$	&	$	\cdots	$	\\	\vspace{1mm}															
1361742	&	15.7609	&	812001	&	9.3982	&	$	\cdots	$	&	$	\cdots	$	&	$	\cdots	$	&	$	-19.89	^{+	0.28	}_{-	0.25	}	$	&	$	\cdots	$	\\	\vspace{1mm}															
1705078	&	19.7347	&	1016731	&	11.7677	&	$	-19.46	\pm	0.04	$	&	$	\cdots	$	&	$	\cdots	$	&	$	\cdots	$	&	$	\cdots	$	\\	\vspace{1mm}																		
1706253	&	19.7483	&	1017432	&	11.7758	&	$	\cdots	$	&	$	-20.06	\pm	0.04	$	&	$	\cdots	$	&	$	\cdots	$	&	$	\cdots	$	\\	\vspace{1mm}																		
1880549	&	21.7656	&	1121363	&	12.9787	&	$	-19.58	\pm	0.15	$	&	$	-20.15	\pm	0.07	$	&	$	-20.31	\pm	0.13	$	&	$	-20.27	^{+	0.19	}_{-	0.18	}	$	&	$	-20.87	^{+	0.39	}_{-	0.35	}	$	\\	\vspace{1mm}				
2049952	&	23.7263	&	1222378	&	14.1479	&	$	\cdots	$	&	$	\cdots	$	&	$	\cdots	$	&	$	\cdots	$	&	$	-20.75	\pm	0.06	$	\\	\vspace{1mm}																		
2401323	&	27.7931	&	1431899	&	16.5729	&	$	-19.52	^{+	0.28	}_{-	0.27	}	$	&	$	-19.80	\pm	0.15	$	&	$	-20.15	\pm	0.17	$	&	$	-20.38	^{+	0.23	}_{-	0.22	}	$	&	$	-20.51	^{+	0.53	}_{-	0.48	}	$	\\	\vspace{1mm}	
2664187	&	30.8355	&	1588644	&	18.3871	&	$	\cdots	$	&	$	\cdots	$	&	$	\cdots	$	&	$	\cdots	$	&	$	-20.43	^{+	0.16	}_{-	0.15	}	$	\\	\vspace{1mm}															
3037306	&	35.1540	&	1811133	&	20.9622	&	$	\cdots	$	&	$	\cdots	$	&	$	\cdots	$	&	$	\cdots	$	&	$	-20.47	^{+	0.22	}_{-	0.21	}	$	\\	\vspace{1mm}															
3085966	&	35.7172	&	1840149	&	21.2980	&	$	\cdots	$	&	$	\cdots	$	&	$	\cdots	$	&	$	\cdots	$	&	$	-20.64	\pm	0.07	$	\\	\vspace{1mm}																		
3090966	&	35.7751	&	1843130	&	21.3325	&	$	-19.17	^{+	0.18	}_{-	0.17	}	$	&	$	-19.84	\pm	0.11	$	&	$	-20.00	^{+	0.17	}_{-	0.16	}	$	&	$	-20.35	\pm	0.19	$	&	$	\cdots	$	\\	\vspace{1mm}						
3518554	&	40.7240	&	2098099	&	24.2836	&	$	\cdots	$	&	$	\cdots	$	&	$	\cdots	$	&	$	\cdots	$	&	$	-20.24	\pm	0.09	$	\\	\vspace{1mm}																		
3692304	&	42.7350	&	2201705	&	25.4827	&	$	\cdots	$	&	$	\cdots	$	&	$	-19.70	\pm	0.12	$	&	$	\cdots	$	&	$	\cdots	$	\\	\vspace{1mm}																		
3693574	&	42.7497	&	2202463	&	25.4915	&	$	\cdots	$	&	$	\cdots	$	&	$	\cdots	$	&	$	-19.84	^{+	0.23	}_{-	0.22	}	$	&	$	\cdots	$	\\	\vspace{1mm}															
3694905	&	42.7651	&	2203256	&	25.5007	&	$	-18.69	\pm	0.07	$	&	$	\cdots	$	&	$	\cdots	$	&	$	\cdots	$	&	$	\cdots	$	\\	\vspace{1mm}																		
3696071	&	42.7786	&	2203952	&	25.5087	&	$	\cdots	$	&	$	-19.45	\pm	0.05	$	&	$	\cdots	$	&	$	\cdots	$	&	$	\cdots	$	\\	\vspace{1mm}																		
3950847	&	45.7274	&	2355874	&	27.2671	&	$	\cdots	$	&	$	\cdots	$	&	$	\cdots	$	&	$	\cdots	$	&	$	-20.24	\pm	0.09	$	\\	\vspace{1mm}																		
4258444	&	49.2875	&	2539292	&	29.3900	&	$	-18.64	^{+	0.39	}_{-	0.37	}	$	&	$	-19.25	\pm	0.20	$	&	$	-19.42	^{+	0.42	}_{-	0.40	}	$	&	$	-19.61	^{+	0.62	}_{-	0.58	}	$	&	$	\cdots	$	\\	\vspace{1mm}			
4732196	&	54.7708	&	2821789	&	32.6596	&	$	-18.36	^{+	0.63	}_{-	0.58	}	$	&	$	-18.77	^{+	0.27	}_{-	0.26	}	$	&	$	-19.25	^{+	0.32	}_{-	0.31	}	$	&	$	-19.38	^{+	0.48	}_{-	0.46	}	$	&	$	\cdots	$	\\	\vspace{1mm}
6241880	&	72.2440	&	3722007	&	43.0788	&	$	\cdots	$	&	$	-17.79	^{+	0.84	}_{-	0.74	}	$	&	$	-18.76	^{+	0.78	}_{-	0.73	}	$	&	$	-18.78	^{+	1.57	}_{-	1.34	}	$	&	$	\cdots	$	\\						
\hline									
\hline									
\end{longtable}
}
}

\subsection{The bolometric light curve of SN 2011kl}
\label{bolometricSN}

\begin{figure*}[t]
  \centering
 \includegraphics[width=\textwidth]{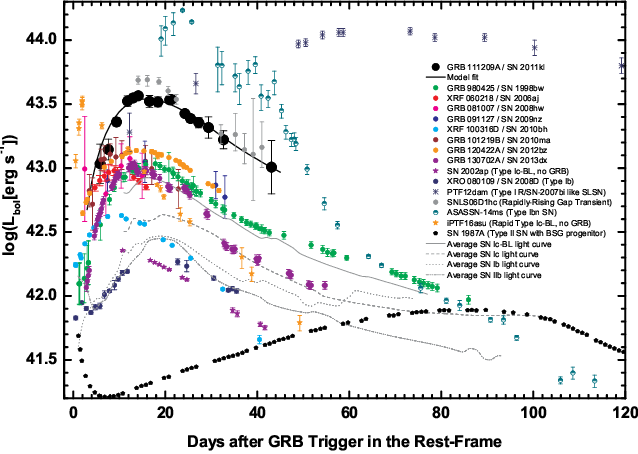}
   \caption{Bolometric light curves of SE-SNe. SN 2011kl is seen to exceed all other well-monitored GRB-SNe \citep{Olivares2012AA,Olivares2015AA,Prentice2016MNRAS} in luminosity, including the bright GRB 120422A/SN 2012bz \citep{Schulze2014AA} and the recent well-monitored GRB 130702A/SN 2013dx \citep{Toy2015ApJ}. It is far less luminous than superluminous supernovae though, exemplified by PTF12dam here \citep{Chen2015MNRAS}, but of similar luminosity as one of the ``rapidly-rising gap transients'' presented by \cite{Arcavi2015ApJ}, the light curve evolution is also similar (we here assign SNLS06D1hc the same peak time as SN 2011kl). {Several further luminous SNe are also shown, see Sect. \ref{RRGTchapter} for more details}. Additionally, we plot mean light curves of four different SE-SNe types taken from \cite{Lyman2014MNRAS} as well as SN 1987A, which had a BSG progenitor \citep{Suntzeff1990AJ}. We assume $t_{peak}=64$ days for PTF12dam \citep{Nicholl2013Natur,Chen2015MNRAS} and $t_{peak}=19$ days for the average curves \citep{Lyman2014MNRAS}.}
              \label{LC_bol}
    \end{figure*}

\begin{table}[t]
\begin{center}
\caption{Results of modelling the SN with a two-component $^{56}$Ni decay model. The expansion velocity is an input parameter.}
\begin{tabular}{lc} 
\hline\hline \vspace{1mm}
Expansion velocity $v_{\rm exp}$ (10$^3$ km s$^{-1}$) & $21\pm7$ \\ \vspace{1mm}
Ejected mass $M_{\rm ej}$ ($M_\odot$) & $6.79^{+3.67}_{-2.84}$ \\ \vspace{1mm}
Ejecta kinetic energy $E_k$ ($10^{51}$ erg) & $34.26^{+34.07}_{-32.00}$ \\ \vspace{1mm}
nickel mass $M_{\rm Ni}$ ($M_\odot$) & $2.27\pm0.64$ \\  \vspace{1mm}
Mass fraction, inner component & $0.31\pm0.19$ \\ \vspace{1mm}
Energy fraction, inner component & $0.000\pm0.003$ \\
\hline\hline
\label{SNExpPar}
\end{tabular}
\end{center}
\end{table}

\onltab{4}{
\longtab{4}{
\begin{longtable}{rc}
\caption{\label{BolLC} The bolometric light curve of GRB-SN 111209A, given in the rest-frame.}\\
\hline 
\hline \vspace{1mm}
$\Delta$t (ks) & Bolometric peak luminosity ($\log(L_{\tn{bol}}) (\tn{erg s}^{-1})$)\\\hline
503	& $	43.03	\pm	0.15	$ \\
657	& $	43.14	\pm	0.08	$ \\
811	& $	43.36	\pm	0.03	$ \\
1017	& $	43.52	\pm	0.04	$ \\
1122	& $	43.55	\pm	0.03	$ \\
1223	& $	43.56	\pm	0.02	$ \\
1432	& $	43.52	\pm	0.05	$ \\
1589	& $	43.52	\pm	0.04	$ \\
1811	& $	43.53	\pm	0.03	$ \\
1842	& $	43.54	\pm	0.03	$ \\
2098	& $	43.43	\pm	0.02	$ \\
2203	& $	43.39	\pm	0.03	$ \\
2356	& $	43.36	\pm	0.04	$ \\
2539	& $	43.32	\pm	0.08	$ \\
2822	& $	43.22	\pm	0.07	$ \\
3722	& $	43.01	\pm	0.21	$ \\
\hline									
\hline									
\end{longtable}
}
}

The bolometric light curve was constructed using the pure SN data from Table 1 in G15 following the methods presented in \cite{Olivares2012AA,Olivares2015AA}. As not all data are contemporaneous, we used a polynomial of second order for interpolation (and a first-order one at late times) to derive a complete SED at each epoch. We fitted the SEDs using quadratic polynomia (Simpson's Rule) and integrated across them (including the observer-frame $J$-band data). These values were then corrected for a time-dependent rest-frame NIR contribution based on data of the GRB-SNe SN 1998bw and SN 2006aj \citep{Olivares2015AA}. The final values represent the bolometric luminosity across the rest-frame $0.3-2.2$ $\mu$m band. 

In Fig. \ref{LC_bol}, we show the bolometric light curve of SN 2011kl (given in Table \ref{BolLC} and peaking at $L_{\rm bol,peak}=(3.63^{+0.17}_{-0.16})\times10^{43}$ erg s$^{-1}$) in comparison to the bolometric light curves of several other GRB- and XRF-SNe, taken from \citet[][their Fig. 7]{Olivares2012AA} as well as \cite{Olivares2015AA}, \cite{Schulze2014AA}, and \cite{Toy2015ApJ}; we additionally include some non-GRB SE-SNe, the SLSN PTF12dam \citep{Chen2015MNRAS}, the ``rapidly-rising gap transient'' SNLS06D1hc \citep{Arcavi2015ApJ} {as well as two more SNe in the ``luminosity gap'' between typical GRB-SNe and SLSNe \citep{Vallely2017MNRAS,Whitesides2017ApJ}}, the famous SN 1987A, which had a Blue SuperGiant (BSG) progenitor \citep{Suntzeff1990AJ}, as well as mean light curves of four different SE-SNe types taken from \cite{Lyman2014MNRAS}. {Note that in some of these cases, only the observer-frame optical range has been used, implying these are pseudo-bolometric light curves and the true luminosities will be higher. This does not change the general picture, though.}

Clearly, SN 2011kl is more luminous than any GRB-SN discovered to date (which tend to cluster in a relatively narrow luminosity range, \citealt{Melandri2014AA}), including the previous record-holder, SN 2012bz associated with GRB 120422A \citep{Schulze2014AA}, and generally the most luminous Type Ic SN which is not a \emph{bona fide} SLSN \citep{Prentice2016MNRAS}. It is also much more luminous and evolves much faster than the BSG-progenitor SN 1987A. On the other hand, it is much less luminous and also much faster evolving than PTF12dam. This SLSN shows a light-curve evolution similar to SN 2007bi, which has been claimed to be due to a very massive star exploding as a pair-instability supernova (PISN, \citealt{Gal-Yam2009Nature}). Such SLSNe have been labelled ``SLSN R'' by \cite{Gal-Yam2012Science}, but \cite{Nicholl2013Natur} and \cite{Chen2015MNRAS} show that other similar SNe, including PTF12dam, can also be powered by magnetars \citep[or possibly black holes,][]{Nicholl2015ApJL} or the interaction of the SN ejecta with a massive shell ejected at an earlier time (the latter model being rather construed, though); therefore they label this SLSN-type ``2007bi-like'' {(note \cite{Nicholl2017ApJ3} and \cite{Decia2017ApJ} find a continuum of rise/decay times from large samples, and no bimodality at all anymore)}. Our main motivation in using this specific SLSN as a comparison in Fig. \ref{LC_bol} is that the {pseudo-}bolometric light curve was freely available. See G15 for a comparison to two faster SLSNe.

Furthermore, we show a fit to the $0.3-2.2$ $\mu$m bolometric light curve using a two-component $^{56}$Ni decay model, with a high-density inner region (which is of high opacity and only dominates in the nebular phase) and a low-density outer region which dominates the early SN emission (based on \citealt{Arnett1982ApJ} and \citealt{Maeda2003ApJ}). As the expansion velocity of the SN could not be determined from spectroscopy because of the lack of any kind of readily detectable absorption lines (L14, G15, \citealt{Mazzali2016MNRAS}), the model assumes a typical (for a Type Ic-BL SN) expansion velocity at peak, namely $v_{\rm exp}=(21\pm7)\times10^4$ km s$^{-1}$ (close to the one used by G15 and \citealt{Mazzali2016MNRAS} for modelling as well). We use a grey opacity of $0.07\pm0.01$ cm$^2$ g$^{-1}$, identical to \cite{Olivares2015AA} and G15. The deduced values of the nickel mass created in the explosion $M_{\rm Ni}$, the ejecta mass $M_{\rm ej}$, the total kinetic energy in the ejecta $E_k$, and the mass and energy fractions of the inner component of the two-component model are given in Table \ref{SNExpPar}. The span of $M_{\rm ej}$ ($3.95-10.46$ $M_\odot$) and $E_k$ ($[2.26-68.33]\times10^{51}$ erg) are large because of the large error in velocity, but we note that the derived nickel mass $M_{\rm Ni}=2.27\pm0.64$ $M_\odot$ is independent of $v_{\rm exp}$.

G15 do not use the $J$-band data, or a NIR correction, and therefore derive lower values of $L_{\rm bol,peak}=(2.8^{+1.2}_{-1.0})\times10^{43}$ erg s$^{-1}$, $M_{\rm ej}=3.2\pm0.5$ $M_\odot$ and $M_{\rm Ni}=1.0\pm0.1$ $M_\odot$. \cite{Cano2016MNRAS}, using values from G15, derive $M_{\rm ej}\approx5.2$ $M_\odot$, in better agreement with our value. \cite{Metzger2015MNRAS} assume that beyond the magnetar heating, a typical amount of nickel is also present ($M_{\rm Ni}\approx0.2$ $M_\odot$), and find that $M_{\rm ej}\approx3$ $M_\odot$, in good agreement with G15, though \cite{Metzger2015MNRAS} use a grey opacity of 0.2 cm$^2$ g$^{-1}$. \cite{Bersten2016ApJL} also model the bolometric data of G15 with their own magnetar model, which uses $M_{\rm Ni}=0.2$ $M_\odot$ and $M_{\rm ej}=2.5$ $M_\odot$, and they also employ a grey opacity of 0.2 cm$^2$ g$^{-1}$. They find a best fit when at least some of the SN heating is due to radioactive decay ($M_{\rm Ni}\geq0.08$ $M_\odot$). \cite{Yu2017ApJ} derive a significantly smaller $M_{\rm ej}=0.51\pm0.06$  $M_\odot$. {Finally, \cite{Wang2017ApJ3} model the bolometric light curve presented in this paper, and find it can also be fit by a pure magnetar model, or a magnetar$+^{56}$Ni model. Using $\kappa=0.07$ cm$^2$ g$^{-1}$, as we do (see also G15), they derive $M_{\rm ej}=4.50^{+1.76}_{-1.16}$ $M_\odot$, $M_{\rm Ni}=0.11^{+0.06}_{-0.07}$ $M_\odot$. Their pure $^{56}$Ni model (again, for $\kappa=0.07$ cm$^2$ g$^{-1}$) results in $M_{\rm ej}=4.57^{+0.80}_{-1.03}$ $M_\odot$, $M_{\rm Ni}=1.42\pm0.04$ $M_\odot$, somewhat lower than our values, but in full agreement with the conclusions of G15 and our work that a pure $^{56}$Ni model is untenable.} As a side note, we point out that using our maximum bolometric luminosity log $L_{peak}=43.56$ and Eq. 2 from \cite{Kozyreva2016MNRAS}, we derive $M_{\rm Ni}=5.6$ $M_\odot$, an even more extreme value. We discuss these results in the context of larger SN samples in Sect. \ref{SectDiscSN}.


\section{Results}
\label{SectRes}

\subsection{Blackbody fit of the pure supernova}
\label{SectBB}

Using the afterglow fit derived in Sect. \ref{secSN} as well as the host galaxy magnitudes, we subtract the individual contributions of afterglow and host for each band, leaving us with the pure magnitudes of the SN. We then also correct these values for the rest-frame extinction derived by K18B. These values are given as Table 1 in G15. Employing our selected cosmology, the redshift $z=0.67702$, and including a correction for the local velocity field \citep{Mould2000ApJ}, we derive a luminosity distance of 4076.5 Mpc, which translates into a distance modulus of $\mu=43.05$ mag. This allows us to translate the tabulated values from G15 into rest-frame time and absolute magnitudes at the rest-frame wavelengths given in Table \ref{tab_fit4}. We give the derived values in Table \ref{SNtabrest}.

\begin{figure}
\includegraphics[width=9.0cm,angle=0]{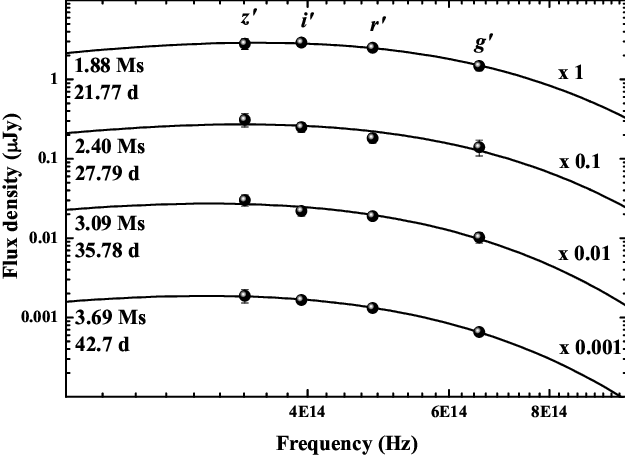}
\caption{Blackbody fits to the GROND $g^\prime r^\prime i^\prime z^\prime $-band data of SN 2011kl (times in observer frame, Table \ref{SNtabrest}). Note that for reasons of clarity the $y$ axis was scaled by the given factors.}
\label{SNbbfig}
\end{figure}

We fitted the flux densities of the supernova component using the {\tt zbbody} tool part of the {\tt Xspec v12.7.1} software package\footnote{\tt http://heasarc.nasa.gov/xanadu/xspec/}, assuming that it can be modelled by blackbody radiation. The quality of our optical photometry $(g^\prime r^\prime i^\prime z^\prime )$ is best for the epochs at 1.88, 2.40, 3.09 and 3.69 Ms/21.77, 27.79, 35.78, and 42.70 days (Table \ref{SNtabrest}). The $J$-band is not included as the flux excess places it above the blackbody fit. In this time span the blackbody temperature $T_{\rm bb}$ (in the host frame) decreased from $9.98^{+0.81}_{-0.70}$ kK to $8.70\pm0.35$ kK. At the same time, the bolometric luminosity $L_{\rm bb}$ dropped from $3.3\,\times$ to $1.8\times10^{43}$ erg/s, while the radius of the emitting shell, defined via $R_{\rm bb}= (L_{\rm bb}/4\pi \sigma T_{\rm bb}^4)^{1/2}$, was about $2.3\times10^{15}$ cm  (see Fig. \ref{SNbbfig}, and Table~\ref{SNbbtab2}).

\begin{table}[t]
\renewcommand{\tabcolsep}{6pt}
\begin{center}
\caption{Results of the blackbody fits.}
\begin{tabular}{clccc} 
\hline\hline   
$\Delta$t (Ms/d) & $T_{\rm bb}$ (kK) & $\chi^2_\nu$ & $L_{\rm bb}$ ($10^{43}$~erg/s) & $R_{\rm bb}$ ($10^{15}$ cm)\\\hline \vspace{1mm}
1.88/21.77    &  $9.98_{-0.70}^{+0.81}$  &  0.04 & $3.27 \pm 0.17$ & $ 2.1 \pm 0.4$ \\ \vspace{1mm}
2.40/27.79    &  $9.17\pm1.28$  &  0.94 & $2.68 \pm 0.24$ & $ 2.3 \pm 0.6$ \\ \vspace{1mm}
3.09/35.78    &  $8.94_{-0.70}^{+0.81}$  &  0.57 & $2.57 \pm 0.20$ & $ 2.4 \pm 0.5$ \\ \vspace{1mm}
3.69/42.70    &  $8.70\pm0.35$  &  0.04 & $1.80 \pm 0.09$ & $ 2.1 \pm 0.2$ \\
\hline\hline
\label{SNbbtab2}
\end{tabular}
\end{center}
\end{table}



\subsection{Absolute magnitudes for SN 2011kl}
\label{Absmags}

Using data from \cite{Clocchiatti2011AJ} and \cite{McKenzie1999PASP}, and correcting for the small Galactic line-of-sight extinction \citep{Schlafly2011ApJ}, we employ the analytical equation derived by \cite{Zeh2004ApJ} to derive the maximum brightness of SN 1998bw in the $U$, $B$, $V$, and $R_C$ bands independent of data scatter and sampling density, finding (Vega magnitudes) $U=13.891$ mag, $B=14.146$ mag, $V=13.673$ mag, and $R_C=13.598$ mag (these values are not yet corrected for host-galaxy extinction). Using the precise host-galaxy redshift of $z=0.00867$ \citep{Foley2006AA}, we derive (using CosmoCalc\footnote{http://www.astro.ucla.edu/$\sim$wright/CosmoCalc.html}, \citealt{Wright2006PASP}, our assumed world model, and the NED Velocity Correction Calculator\footnote{http://ned.ipac.caltech.edu/forms/vel\_correction.html}) a luminosity distance of $D_L=35.5$ Mpc to SN 1998bw. From this, we derive a distance modulus of 32.75 mag, and thus $M_U=-19.17$ mag, $M_B=-18.87$ mag, $M_V=-19.28$ mag, and $M_{R_C}=-19.31$ mag for SN 1998bw. Here, we assume a host-galaxy extinction for SN 1998bw of $A_V=0.2$ mag \citep{Woosley1999ApJ}, the same as was used to derive the template light curves for all our GRB-SN fits. This value is fully in agreement with the result derived at the SN location by \cite{Kruhler2017AA}.

At the redshift of GRB 111209A/SN 2011kl, the observed filters do not correspond to the rest-frame $UBVR_C$ filters, although $U_{rest}$ is essentially the same as $r^\prime$, $i^\prime$ lies just slightly redward of $B_{rest}$, and $z^\prime$ lies slightly blueward of $V_{rest}$, in terms of central wavelengths. Only $R_{C,rest}$ lies roughly in the middle between $z^\prime$ and $J$, corresponding to $Y$. We find that the five $k$ values we derive, if plotted over the observer-frame central wavelengths of the filters, can be well-interpolated ($\chi^2=0.89$ for 2 degrees of freedom) with a polynomial of second order. From this fit, we derive $k_{U,rest}=3.09\pm0.16$, $k_{B,rest}=2.05\pm0.13$, $k_{V,rest}=1.62\pm0.15$, and $k_{R_C,rest}=2.33\pm0.18$, leading to $M_U=-20.39\pm0.06$ mag, $M_B=-19.65\pm0.07$ mag, $M_V=-19.80\pm0.10$ mag, and $M_{R_C}=-20.23\pm0.09$ mag. Note that this fit ignores the spectral dissimilarities between SN 1998bw and SN 2011kl (G15), but as mentioned most rest-frame central wavelengths lie close to observer-frame bands.


\section{Discussion}
\label{SectDisc}

\subsection{The SN associated with GRB 111209A in the context of other GRB-SNe, SE-SNe, and SLSNe}
\label{SectDiscSN}

The excellent agreement with the temporal evolution of SN 1998bw (Sect. \ref{secSN}) gives us high confidence that we are indeed seeing a SN following GRB 111209A, one that is significantly more luminous than the prototypical SN 1998bw itself, indeed, it is the most luminous GRB-SN (with a high-confidence detection) found so far. This conclusion is fully borne out by the spectroscopic classification of the SN (G15). Relativistic tidal disruption flares \citep{Levan2011Science,Cenko2012ApJ,Brown2015MNRAS} are not expected to be accompanied by any kind of significantly brightening late-time emission which would look similar to a {SN.}
This is a further indicator that GRB 111209A is an extreme case of a classical GRB.

While there is strong evidence that SN 2011kl is not fully powered by radioactive decay (G15, \citealt{Metzger2015MNRAS,Bersten2016ApJL,Cano2016MNRAS,Yu2017ApJ,Wang2017ApJ3}), we will for now continue with the results from the $^{56}$Ni modelling, to show just how much of an outlier SN 2011kl is in such a context.

To allow a direct comparison of SN 2011kl and other GRB-SNe, we have also undertaken systematic multi-colour afterglow/SN analysis of a large sample of GRB-SNe (Kann et al., in prep.) . The main conclusion of this analysis is twofold: several of the multi-colour GRB-SNe show divergences from SN 1998bw spectrally (e.g., SNe 2010bh and 2012bz), and none exceed SN 2011kl in luminosity. The exceptional ultraviolet luminosity of SN 2011kl therefore seems unique, at least in comparison to the currently known sample.

\subsubsection{SN 2011kl and superluminous SNe}

In recent years, a special class of ultra-luminous, UV-bright (e.g., \citealt{Tolstov2017ApJ2}) transients has been recognized by \citet[][see \citealt{Gal-Yam2012Science} {and \citealt{Moriya2018SSRv,Gal-Yam2019ARAA} for reviews}]{Quimby2011Nature}. \cite{Pastorello2010ApJ}, presenting more detailed observations of one of the \cite{Quimby2011Nature} sources, link these to Type Ic SNe, implying that UV-suppression is not a given for such SE-SNe, at least not around peak time \citep[see also][]{Inserra2013ApJ}. Indeed, nebular spectroscopy of several events reveals that at very late times, they are spectroscopically indistinguishable from SNe associated with {GRBs \citep{Nicholl2016ApJ2,Jerkstrand2017ApJ,Kangas2017MNRAS,Quimby2018ApJ}.}
The classical, albeit arbitrary definition of SLSNe is $M_U<-21$ at peak, therefore SN 2011kl is not a SLSN per se. Recently, though, the designation has been applied to other luminous SNe that fall under the luminosity limit but are spectrally clearly similar to SLSNe \citep[e.g.,][]{Decia2017ApJ,Lunnan2018ApJ,Quimby2018ApJ,Angus2019MNRAS} leading other authors to designate SN 2011kl as a SLSN \citep[e.g.,][]{Liu2017ApJ,Margutti2017ApJX}.

{Within the statistical ``Four Observables Parameter Space (4OPS)'' context \citep[see][for details]{Inserra2018ApJ4OPS}, we find that SN 2011kl is not fully but mostly in agreement with the parameter space of SLSNe, in contrast to what \cite{Inserra2018ApJ4OPS} find for GRB-SNe. We note, however, that \cite{Angus2019MNRAS} have also presented multiple spectroscopically classified SLSNe that do not agree with all four panels. It is therefore unclear how strong a diagnostic tool this is at this stage.}

{It has been found that these SLSNe are also found in dwarf host galaxies which seemed to resemble those of GRBs (\citealt{Chen2013ApJ}, \citealt{Lunnan2013ApJ,Lunnan2014ApJ}), but have been shown to have even more extreme properties (e.g., \citealt{Leloudas2015MNRAS,Thoene2015MNRAS,Perley2016ApJ,Schulze2017MNRAS}).
The single exception may be the host of the closest SLSN, SN 2017egm, which is a large spiral galaxy for which {an $\approx$solar or even super-solar} metallicity is claimed \citep{Nicholl2017ApJ2,Bose2017ApJ,Chen2017ApJ2,Yan2018ApJ}, but see \cite{Izzo2017AA} concerning metallicity diagnostics{, these authors find $Z=0.6Z_\odot$, just above the possible metallicity cutoff \citep[e.g.][]{Schulze2017MNRAS}}.}
Indeed, SLSNe may be the very first SNe to occur in the youngest starbursts, even earlier than GRBs \citep{Leloudas2015MNRAS,Thoene2015MNRAS}. Similar to GRBs, they can be detected to very high redshifts, {ranging from $z\approx1-4$ \citep{Cooke2012Nature,
Moriya2018ApJ,Curtin2018ApJ,Angus2019MNRAS}}. While \cite{Moriya2010ApJ} suggest it may be possible, none of these events have been associated with GRBs or relativistic blastwaves in general \citep{Coppejans2018ApJ}. \cite{Sanders2012ApJ} specifically look for a connection in the case of SN 2010ay, which strongly resembles GRB-SNe, but, excepting that event, these transients are spectroscopically quite different from GRB-SNe. {In contrast to GRBs and their afterglows/SNe, t}hey have also not yet been detected at very high energies (\citealt{Renault-Tinacci2017AA}
), in gamma-rays, X-rays (\citealt{Margutti2017ApJX,Bhirombhakdi2018ApJ}, with a few exceptions, \citealt{Levan2013ApJ,Margutti2017ApJX}), or at radio wavelengths (\citealt{Coppejans2018ApJ}
).

While being comparatively UV-luminous, SN 2011kl does not exhibit such high blackbody temperatures as are found for these SLSNe around maximum, which are typically in the range of $14-17$ kK \citep[e.g.,][]{Quimby2013MNRAS}.
Some SLSNe have lower temperatures at peak which are more comparable to that of SN 2011kl, {in the range $10-12$ kK, \citep[e.g.,][]{McCrum2013MNRAS,McCrum2015MNRAS,Howell2013ApJ}},
and even lower values have been measured, e.g., iPTF13ehe at 7 kK \citep{Yan2015ApJ}. SN 2011kl is hotter than usual Type Ic-BL SNe, though; several GRB-SNe shown in \citet[][their Figure 18]{Nicholl2015MNRAS} show temperatures at peak of $6-8$ kK.

In terms of luminosity, SN 2011kl falls below the luminosities of most SLSNe. A direct comparison free of any bolometric transformations can be done vs. the SLSN DES13S2cmm, which lies at almost exactly the same redshift \citep{Papadopoulos2015MNRAS}, this SLSN is about 0.9 mag brighter at peak than SN 2011kl. The largest $L_{\rm bb}$ we measure is $(3.67\pm0.21)\times10^{43}$ erg s$^{-1}$, while SLSNe may reach up to $>200\times10^{43}$ erg s$^{-1}$ (ASASSN-15lh, \citealt{Dong2015Sci,Godoy-Rivera2017MNRAS,Brown2016ApJ}, but see \citealt{Leloudas2016NatAst,Margutti2017ApJ2,Kruehler2017AA2}, who present strong evidence that this is actually a tidal disruption event). Figure \ref{SLSNeBol} shows\footnote{See Appendix for the data sources.}, though, that not only is it more luminous at peak than all known GRB-SNe, it also is comparable to the least luminous of the SNe that have been labelled superluminous, at least {pseudo-}bolometrically. We caution here that while all GRB-SNe shown except for SN 2016jca \citep{Ashall2017Nature,Cano2017AA}{, iPTF17cw \citep{Corsi2017ApJ}, SN 2017htp \citep{deUgartePostigo2017GCN}{, and SN 2017iuk \citep{Izzo2019Nature}}} have had their bolometric luminosities determined with the same method \citep{Olivares2015AA}, the bolometric luminosities of the SLSNe and other transients are usually based on observer-frame optical data only {(pseudo-bolometric)} and therefore their true bolometric luminosities may be higher.

\begin{figure}[t!]
  \centering
 \includegraphics[width=\columnwidth]{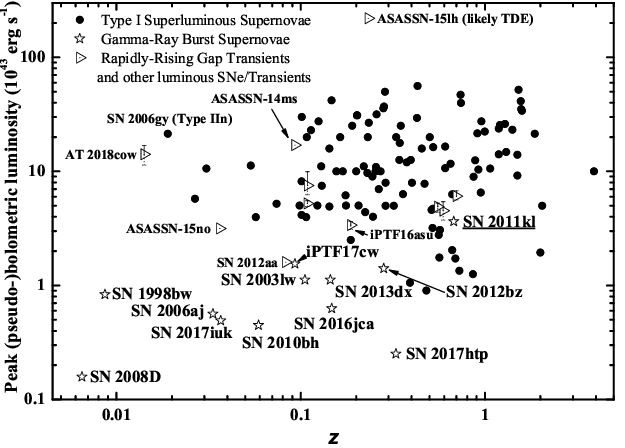}
   \caption{A comparison between the bolometric luminosity at peak of GRB-SNe, SLSNe{, the ``rapidly-rising gap transients'' \citep[RRGTs, ][]{Arcavi2015ApJ} and other luminous SNe (see Sect. \ref{RRGTchapter} for more details}). SN 2011kl is more luminous than any GRB-SN known, and comparable to the least luminous SLSNe {(being possibly even more luminous than a few)} and the RRGTs. Note here that while most of the GRB-SNe have bolometric luminosities that include a NIR correction \citep{Olivares2012AA,Olivares2015AA}, this is not the case for all SLSNe or the RRGTs, therefore in some cases, the bolometric luminosities may be underestimated. Furthermore, SN 2008D, a Type Ib SN, is not associated with relativistic ejecta \citep{Soderberg2008Nature,Malesani2009ApJ} and therefore strictly not a GRB-SN, but a possible ``transition object''.}
              \label{SLSNeBol}
    \end{figure}

{It should be shortly mentioned that pair-instability SNe \citep{Barkat1967PhRvL, Fraley1968Ap&SS}, which have been posited as viable SLSN progenitors \citep[e.g.,][but see, e.g., \citealt{Nicholl2013Natur}]{Gal-Yam2009Nature,Cooke2012Nature}, are very unlikely to be linked to GRBs, even ultra-long ones, as they disrupt the star entirely, leaving no compact remnant. Furthermore, the host galaxy of GRB 111209A is also likely too metal-enriched to host such super-massive stars (L14, \citealt{Stratta2013ApJ}; despite having a low metallicity within the ensemble of GRB host galaxies, \citealt{Kruhler2015AA}). On the other hand, the magnetar model which we favour for SN 2011kl (G15) has been found to be able to fit SLSNe well, both photometrically and spectroscopically \citep[e.g.,][]{Nicholl2013Natur,Nicholl2017ApJ3,Mazzali2016MNRAS}.}

\begin{figure}[t!]
  \centering
 \includegraphics[width=\columnwidth]{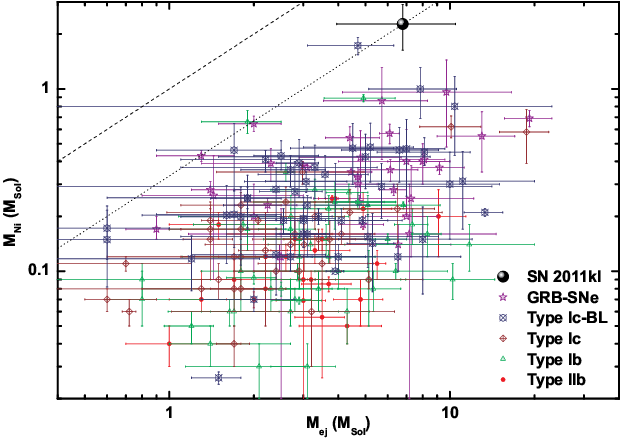}
 \includegraphics[width=\columnwidth]{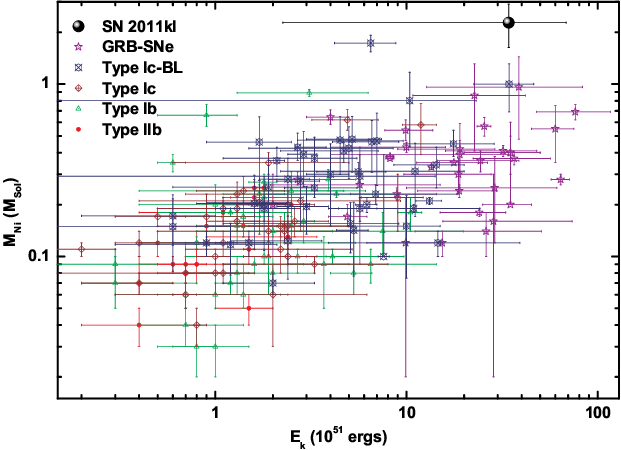}
 \includegraphics[width=\columnwidth]{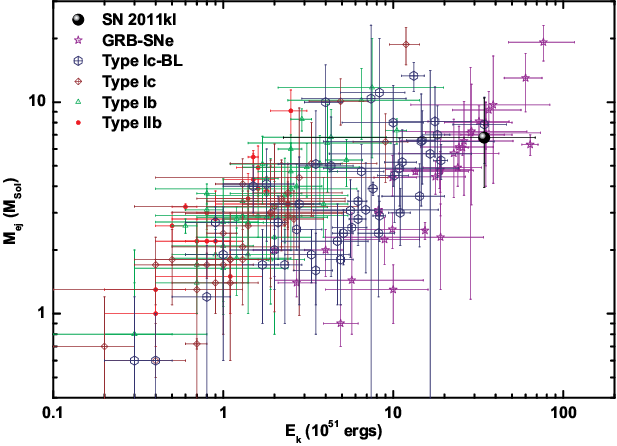}
   \caption{The explosion parameters of SN 2011kl in the context of SE-SNe following \cite{Lyman2014MNRAS}. \textbf{Top panel}: The created $^{56}$Ni mass $M_{\rm Ni}$ vs. the ejected shell mass $M_{\rm ej}$ (both $M_{\odot}$). SN 2011kl has the highest $^{56}$Ni mass of any SN in all the samples, but its ejecta mass is not remarkable. We also plot the extremal case $M_{\rm Ni}=M_{\rm ej}$ (dashed line) as well as $M_{\rm Ni}=M_{\rm ej}/3$ (dotted line), the case for SN 2011kl. \textbf{Middle panel}: $M_{\rm Ni}$ vs. the kinetic energy in the ejecta $E_k$ (in $10^{51}$ erg). SN 2011kl is once again extreme in terms of $M_{\rm Ni}$, but its $E_k$ is comparable to most other GRB-SNe. \textbf{Bottom panel}: $M_{\rm ej}$ vs. $E_k$. Here SN 2011kl is comparable to the rest of the GRB-SN sample.}
              \label{ExpPar}
    \end{figure}

\cite{Nicholl2015MNRAS} studied a moderately large sample of SLSNe. They derived rise and decline times for the pseudo-bolometric light curves, defined as the time between when the luminosity is at $L_{peak}/e$ to $t_{peak}$, and from then to when it has declined again to $L_{peak}/e$. We follow their method and fit our bolometric light curve with a fourth-order polynomial, finding $t_{peak}=16.1$ days in the rest-frame, log $L_{peak}=43.58$, and rise and decline times of 9.1 and 21.1 days, respectively, from this fit. This implies the light curve is narrower than those of all SLSNe in the sample of \cite{Nicholl2015MNRAS}, not just the SLSN-2007bi-like SLSNe as stated above (see also G15). We also compared SN 2011kl with the (i)PTF sample recently presented by \cite{Decia2017ApJ}, who, instead of using $e$, derive rise and decline times for 1 mag and a factor 2. For these values, we find rise times of 8.8 and 7.8 days, and decline times of 19.2 and 14.4 days, respectively. Only PTF 09as (11/9 days) and PTF 10aagc (14/10 days) decline faster, and iPTF13bjz rises comparably fast (9/8 days). Interestingly, the two former SLSNe are among the least luminous in their sample, comparable to or even fainter than SN 2011kl. iPTF 13bzj is more luminous, but still lies beneath the classic SLSN luminosity cutoff, and is sparsely sampled.

SN 2011kl agrees well with the full-sample correlation of \cite{Nicholl2015MNRAS} (their Figure 4), but also does not deviate significantly from the Type Ibc SNe in this plot. Furthermore, \cite{Nicholl2015MNRAS} discussed whether, in luminosity space, there are two distinct populations of Type Ic SNe (normal + BL vs. SLSNe) or whether there is a continuum. {Indeed, SN 2011kl is a transition object indicating that a continuum indeed exists (similar to what has been found for hydrogen-rich SLSNe, \citealt{Inserra2018MNRAS}), see also G15 and the discussion below (Sect. \ref{RRGTchapter})}.

In the context of the magnetar model, \cite{Yu2017ApJ} study a large sample of SLSNe bolometric light curves (they include SN 2011kl within this sample). Within the spread of values they derive, SN 2011kl is seen as an outlier. It is the least luminous SLSN in the sample, it has the lowest ejecta mass (smaller by an order of magnitude compared to our $^{56}$Ni modelling), longest initial magnetar spin period, and lowest rotational energy, as well as one of the lowest spin-down luminosities and one of the strongest magnetic fields (\citealt{LiuLD2017ApJ} also model SLSNe (but not SN 2011kl) with a magnetar model, and all their SLSNe have more rapidly spinning magnetars and larger ejecta masses than SN 2011kl as derived by \citealt{Yu2017ApJ}). The light curve also shows one of the fastest rise times. Only the light-curve decay time as well as the spin-down timescale of the magnetar are found to be average values within the distributions. But in the correlations that \cite{Yu2017ApJ} study, SN 2011kl is found to be at the extreme end in several cases; while still agreeing with the trends, it is not an outlier. They also compare their SLSN sample with values derived for GRB-SNe by \citet[][here, caution must be exercised as the magnetars in their modelling also power the GRB prompt and afterglow emission, and not just the SNe]{Lue2014ApJ}, and it can be seen that in terms of magnetar field strength, SN 2011kl lies at the boundary between SLSNe (low field strengths) and GRB-SNe (high field strengths) -- once more a transition object (it is not unique in this aspect, though). They note that the magnetic field strength is strong enough to launch a jet, but it is near the critical field strength (for GRB-SNe, the field is always strong enough, for SLSNe, the field in most cases is \emph{not} strong enough).

\subsubsection{SN 2011kl and SE-SNe}

To put SN 2011kl into a larger context of more similar SNe, we look into the literature for large comparison samples of SE-SNe (Types IIb, Ib, Ic, BL-Ic, GRB-SN).

\cite{Cano2013MNRAS} used a bolometric approximation method to derive bolometric properties for both GRB/XRF SNe as well as non-GRB-related Type Ibc SNe (some of them BL). In an older paper \citep{Cano2011MNRAS}, they also presented a sample of absolute magnitudes $M_V$ for Type Ibc SNe. \cite{Cano2016JAA} derived improved ``average'' GRB-SN values based on updated data. \cite{Richardson2014AJ} studied large samples of all types of SNe (including non-SE SNe and Type Ia SNe) and derived median absolute $M_B$ magnitudes for the different samples. They find $\overline{M}_B=-16.99\pm0.45$ mag, $\overline{M}_B=-17.45\pm0.33$ mag, and $\overline{M}_B=-17.66\pm0.40$ mag for Type IIb, Type Ib, and Type Ic SNe, respectively, far fainter than what we find for SN 2011kl. \cite{Lyman2014MNRAS} derived bolometric light curves for 38 SE-SNe and used them to determine explosion parameters; this sample partially overlaps with that of \cite{Cano2013MNRAS}. Finally, \cite{Walker2014MNRAS} compiled a literature sample of explosion parameters for Type Ic-BL SNe as well as presenting their analysis of PTF10qts. We take four Type Ic-BL SNe from this paper which are not found in the other samples.

\cite{Cano2013MNRAS} have found that in terms of $M_{\rm Ni}$, $M_{\rm ej}$ and $E_k$, the SNe associated with GRBs and XRFs yield significantly higher values than those derived for ``normal'' Type Ib/c SNe, and even for Type Ic-BL SNe not associated with GRBs/XRFs. For GRB/XRF SNe, the median values they derived are: $\overline{M_{\rm Ni}}=0.3-0.35$ $M_\odot$, $\overline{M_{\rm ej}}=6.0$ $M_\odot$ and $\overline{E_k}=20\times10^{51}$ erg. The values we derive for SN 2011kl (Sect. \ref{bolometricSN} and Table \ref{SNExpPar}) are comparable in terms of $M_{\rm ej}$ (indeed, \cite{Lyman2014MNRAS} find that $M_{\rm ej}$ is the one parameter which is distributed evenly among all the SE-SNe classes), for the most part higher in terms of $E_k$, and significantly higher in terms of $M_{\rm Ni}$. This result remains unchanged also in comparison to the ``average'' GRB-SN derived by \cite{Cano2016JAA}, who found (excluding SN 2011kl itself): $\overline{L_p}=(1.03\pm0.36)\times10^{43}$ erg s$^{-1}$ (28\% of SN 2011kl); $\overline{E_k}=(25.2\pm17.9)\times10^{51}$ erg (74\% of SN 2011kl, identical within errors); $\overline{M_{\rm ej}}=5.9\pm3.8$ $M_\odot$ (87\% of SN 2011kl, identical within errors); $\overline{M_{\rm Ni}}=0.37\pm0.20$ $M_\odot$ (16\% of SN 2011kl). The two highest $^{56}$Ni masses \cite{Cano2013MNRAS} found for GRB-SNe are both also very uncertain, $M_{\rm Ni}\approx0.9\pm0.5$ $M_\odot$ for the SNe accompanying GRBs 991208 and 080319B (see also \citealt{Cano2016JAA}). On the opposite end, $^{56}$Ni masses can go down to just $\approx5$\% of what we find for SN 2011kl (GRB 060904B, XRF 100316D/SN 2010bh, \citealt{Olivares2012AA}, \citealt{Cano2013MNRAS}, \citealt{Cano2016JAA}). Furthermore, none of the $^{56}$Ni masses for any other Type Ibc SN come close to what we find for SN 2011kl. This is also true if the {other samples are} taken into account. Such a $^{56}$Ni mass also far exceeds what could be produced by a magnetar alone (\citealt{Suwa2015MNRAS}, \citealt{Chen2017ApJ}){, although \cite{Song2019ApJ} claim it can be produced in the outflow for low-metallicity progenitors}. 

\cite{Lyman2014MNRAS} presented a large sample of bolometric light curves (their Fig. 1). Using their conversion between peak bolometric luminosity and absolute bolometric magnitude, and using the peak bolometric luminosity we derive for SN 2011kl ($\log L_{\rm bol}=43.56$), {we find that at peak, $M_{\rm bol}=-20.20\pm0.05$ mag. Using our SN 2011kl values and equation (4) of \cite{Lyman2014MNRAS}, we either {find} $M_{\rm Ni}\approx1.6$ $M_\odot$ by using $M_{\rm bol}$}, or we conversely use our modelled value of $M_{\rm Ni}$ to predict $M_{\rm bol}\approx-20.58$ mag. These values are in reasonable agreement
with the relation from \cite{Lyman2014MNRAS}. They also plotted different explosion parameters they had derived versus each other, we follow their methodology and put SN 2011kl into this context, using an expanded sample (Fig. \ref{ExpPar}) with additional results from sample papers: \cite{Cano2013MNRAS}, \cite{Taddia2018AA,Taddia2019AA2}, \citet[][only $M_{Ni}$ and $M_{ej}$]{Prentice2019MNRAS}, as well as several single events: \citet[][GRB 130702A/SN 2013dx]{D'Elia2015AA}, \citet[][GRB 130702A/SN 2013dx]{Toy2015ApJ}, \citet[][iPTF17cw]{Corsi2017ApJ}, and \citet[][GRB 171205A/SN 2017iuk]{Izzo2019Nature}. Similar to the averaged-values comparison, we find SN 2011kl to be exceptional in terms of $M_{\rm Ni}$, but ordinary in terms of $M_{\rm ej}$ and $E_k$. In the top-most panel, we plot the extremal case $M_{\rm Ni}=M_{\rm ej}$ as well as $M_{\rm Ni}=M_{\rm ej}/3$, the case for SN 2011kl and four other SNe: SN 2005hg, SN 2009ca, SN 2010ma, and iPTF17cw. The first two are the most luminous SNe in \cite{Lyman2014MNRAS} and \cite{Taddia2018AA}, respectively, SN 2010ma is associated with GRB 101219B, while iPTF17cw is an engine-driven Type Ic-BL SN possibly associated with a GRB. This may imply an additional heating source beyond radioactive heating for {these SNe} as well. Note that SNe exhibiting $M_{\rm ej}/E_k= \tn{const.}$ in the third panel are from \cite{Cano2013MNRAS} and the seeming correlations stem from their analysis method, i.e., an assumed constant peak photospheric velocity for those SNe that did not have a this value spectroscopically measured.

\begin{figure}[t!]
  \centering
 \includegraphics[width=\columnwidth]{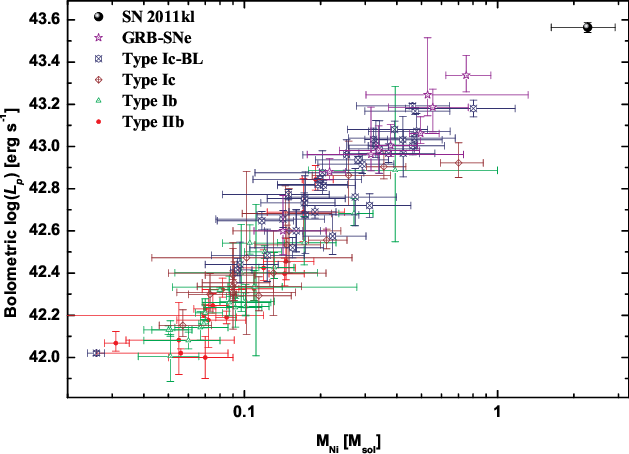}
 \includegraphics[width=\columnwidth]{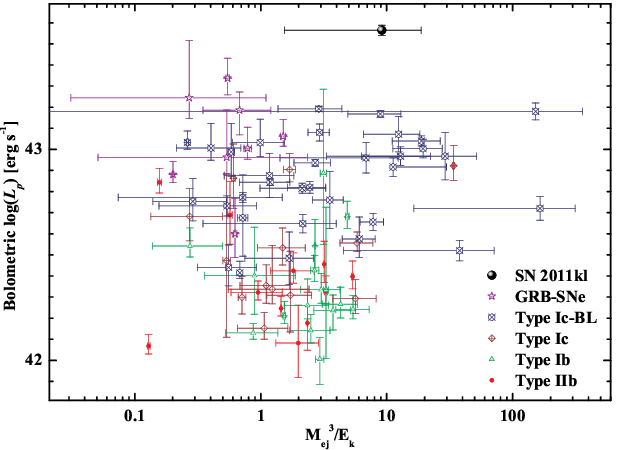}
 \includegraphics[width=\columnwidth]{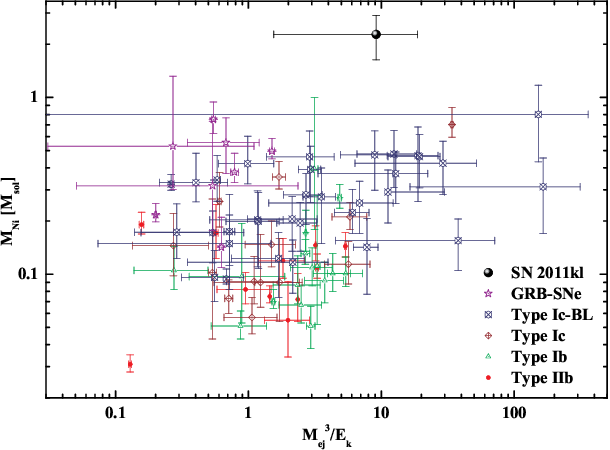}
   \caption{The explosion parameters of SN 2011kl in the context of SE-SNe following \cite{Prentice2016MNRAS}. \textbf{Top panel}: The logarithmic bolometric peak luminosity $L_{peak}$ vs. the created $^{56}$Ni mass $M_{\rm Ni}$ ($M_{\odot}$).  \textbf{Middle panel}: $L_{peak}$ vs. the parameter $M_{\rm ej}^3/E_k$ derived from the ejecta mass $M_{\rm ej}$ and the kinetic energy $E_k$. \textbf{Bottom panel}: $M_{\rm Ni}$ vs. $M_{\rm ej}^3/E_k$. SN 2011kl is an extreme event in all three plots.}
              \label{ExpPar2}
    \end{figure}

\cite{Prentice2016MNRAS} have presented {one of the largest samples of non-SLSN SE-SNe so far, over 80, which they analyse consistently}. SN 2011kl is the most luminous SN in the sample (see their Figs. 8 and 12). They derived log $L_{peak}=43.529^{+0.174}_{-0.148}$ for SN 2011kl (in excellent agreement with our own value), which exceeds the median value they found for the fully bolometric Type Ic-BL/GRB-SNe sample by 0.72 dex. None of their $M_{\rm Ni}$ values exceed $M_{\rm Ni}\approx0.8$ M$_\odot$ (they did not derive $M_{\rm Ni}$ for SN 2011kl itself, simply stating that it is magnetar-powered), and they found a median for the above-mentioned sample of $M_{\rm Ni}=0.34^{+0.13}_{-0.19}$ $M_\odot$, far below our SN 2011kl result. The median values of log $L_{peak}$ and $M_{\rm Ni}$ for all the other SE-SNe classes (non-BL Ic, Ib, IIb) are yet again lower. We adopt their Figs. 19, 20, and 21, and show SN 2011kl in comparison to their sample (Fig. \ref{ExpPar2}, we also add the samples of \citealt{Prentice2019MNRAS} (top panel only) and \citealt{Taddia2019AA2}). The errors we find for $M_{\rm ej}^3/E_k$ for SN 2011kl are large, $M_{\rm ej}^3/E_k=9.15^{+9.6}_{-7.6}$. Clearly, SN 2011kl is a strong outlier. In the top plot of Fig. \ref{ExpPar2}, there is a tight correlation between log $L_p$ and $M_{\rm Ni}$, as expected. SN 2011kl lies beyond all SNe from \cite{Prentice2016MNRAS,Prentice2019MNRAS,Taddia2019AA2} but is in agreement with the correlation. This correlation implies that the middle and bottom plots contain essentially the same information. The value we derive for $M_{\rm ej}^3/E_k$ for SN 2011kl is the second largest in the entire sample, but not extreme. There is a rough trend visible of decreasing $M_{\rm Ni}$ (or log $L_p$) with increasing $M_{\rm ej}^3/E_k$, and SN 2011kl lies outside the main ``cloud'' -- but so do two \cite{Prentice2016MNRAS} events {and multiple of the Type Ic-BL SNe studied by \cite{Taddia2019AA2}}.

All in all, we find that if we model SN 2011kl as a purely $^{56}$Ni-decay powered SN, we find a good fit with a model, but derive results which are physically unrealistic, especially in the context of the spectrum/SED of the SN \citep[see G15,][for details]{Mazzali2016MNRAS}. This remains true even in the case of not using the NIR bolometric correction, which leads to a halved $^{56}$Ni mass compared to our result (G15).

\subsubsection{Powering SN 2011kl}
\label{Power}
An alternative way to power a luminous SN is the interaction with a dense stellar wind, or ejected shells of large mass, the CSM-interaction model. Such shells can be emitted shortly (on astronomical timescales) before the final core collapse, and cause spectra to change from Type I to Type II at late times (e.g., SN 2014C, \citealt{Milisavljevic2015ApJ2}{;
SN 2004dk, \citealt{Mauerhan2018MNRAS}; SN 2017ens, \citealt{Chen2018ApJL},} and SN 2017dio, \citealt{Kuncarayakti2018ApJ}), and possibly even serve as an additional energy source powering SLSNe (e.g., the cases of iPTF13ehe, iPTF15esb, and iPTF16bad; \citealt{Yan2015ApJ,Yan2017ApJ2,Liu2018ApJ2}, \citealt{Wang2016ApJ}{).}
The possibility of such an interaction without the usual spectral signatures of H and He, which are clearly not detected (L14, G15, \citealt{Kruhler2015AA}, \citealt{Mazzali2016MNRAS}), has been discussed in the literature {\citep[e.g.,][]{Blinnikov2010arXiv,Chevalier2011ApJL,Chatzopoulos2012ApJ}}, 
and observed in at least one {case \citep{Ben-Ami2014ApJ},}
but these models have not been discussed in the context of GRB progenitors and have multiple issues when compared with our light-curve evolution. Furthermore, the detailed spectroscopic modelling strongly rules out any sign of interaction \citep[G15,][]{Mazzali2016MNRAS}. In G15, we arrived at the conclusion that SN 2011kl is a lower-luminosity pendant to SLSNe powered by magnetar energy injection, implying that the central engine of GRB 111209A was itself a rapidly spinning magnetar. Note that one proposed indicator for a magnetar central engine, an early shock breakout \citep{Kasen2015ApJ}, would be hidden beneath the bright afterglow which is much more luminous than SN 2011kl even at peak. Further analysis of our data by other teams has yielded results which are in agreement with ours \citep{Metzger2015MNRAS,Bersten2016ApJL,Liu2017ApJ,Cano2016MNRAS,Wang2017ApJ3}, with \cite{Cano2016MNRAS} even stating that the afterglow of GRB 111209A can be powered by magnetar emission (they only use the late afterglow data presented in G15, though, similar to \citealt{Gompertz2017ApJ}). Note that recently it has been proposed, in contrast to the main conclusion of \cite{Cano2016MNRAS}, that many Type Ic-BL SNe may be mostly powered by magnetars \citep{WangLJ2016ApJ,Wang2017ApJ2,Wang2017ApJ1}{, see \cite{Taddia2019AA1} for an observational example}.

\subsubsection{SN 2011kl, Rapidly-Rising Gap Transients and other luminous SNe}
\label{RRGTchapter}

\cite{Arcavi2015ApJ} have recently presented the discovery of several luminous transients, one of which has been spectroscopically confirmed as a SN. The other three are very likely SNe as well. They label these sources ``rapidly-rising gap transients'' (RRGTs), as they rise rapidly to peak luminosity (especially compared with SLSNe) and their peak luminosities are situated between those of usual SE-SNe (and even SNe Type Ia) and SLSNe, a parameter space where few events are known so far (\citealt{Moriya2016ApJ} present a model where a supramassive NS collapses to a BH before expending much of its rotational energy, explaining the lower luminosity and more rapid late decline compared to SLSNe). They draw a link to the similarly luminous and distant SN 2011kl (see Fig. \ref{SLSNeBol}) which would be the first reported RRGT if these events are indeed a distinct class. One source, SNLS06D1hc, stands out especially as its light-curve evolution is very similar to that of SN 2011kl (Fig. \ref{LC_bol}). We refer to \cite{Arcavi2015ApJ} for an in-depth discussion on how the RRGTs compare to SN 2011kl (based partially on the bolometric values from G15), but make a few points here:
The single source which is a spectroscopically confirmed SN, PTF10iam, is formally classified a Type II SN based on a high-velocity feature interpreted as H$\alpha$. G15 find no evidence for H and He in the spectrum of SN 2011kl, and the H emission lines can be explained completely by host-galaxy emission linked to star-formation \citep{Kruhler2015AA}.
\cite{Arcavi2015ApJ} find no GRBs linked to these events, though they only performed a catalogue search. GRB 111209A, while very energetic, had a low peak luminosity and did not trigger Konus-\emph{WIND} \citep{Golenetskii2011GCN}. Had not \emph{Swift} triggered on the GRB, it is unclear when it would have been discovered (see the appendix of K18B for several untriggered extremely long GRBs which were only detected much later by manual inspection). Therefore, such an archive inspection may yield detections of low-luminosity EL-GRBs, or at least stricter upper limits. GRB 111209A has one of the largest isotropic energy releases among $z<0.9$ GRBs and may therefore be a ``bright'' outlier.
The most similar event, SNLS06D1hc, has a single spectroscopic observation which yielded no significant SN features, but there may be a blue continuum superimposed upon the host-galaxy spectrum. The spectrum of SN 2011kl would have been very hard to disentangle from the spectrum of its host galaxy if the host had had a similar luminosity as that of SNLS06D1hc, therefore, we may be seeing a SN 2011kl-like spectrum here as well.
The hosts of the three RRGTs except SNLS06D1hc show roughly solar metallicities, much higher than in the case of SN 2011kl. The metallicity of the host of SNLS06D1hc could not be measured, but the SFR is low, again in contrast to the host of GRB 111209A/SN 2011kl. All RRGT hosts also show evidence of evolved stellar populations.

In recent years, large surveys have begun to fill the ``gap''. Aside from the fainter SLSNe mentioned above, a heterogenous population is being revealed. To just mention some examples: The slowly evolving Type Ic SN 2012aa \citep{Roy2016AA}, the rapidly evolving, very blue Type Ibn ASASSN-14ms \citep{Vallely2017MNRAS}, the extremely fast broad-lined Type Ic-BL iPTF16asu \citep{Whitesides2017ApJ}, and the spectroscopically evolving ASASSN-15no \citep{Benetti2018MNRAS}.

These transients as well as the ones \cite{Arcavi2015ApJ} have detected are clearly rare, and there are some intriguing similarities to SN 2011kl, just as there are clear differences, especially concerning the progenitor environments. Therefore, it {already seems erroneous} to lump all these transients into a separate class. Upcoming high-etendue optical surveys such as the Zwicky Transient Facility \citep{Smith2014SPIE} and the Large Synoptic Survey Telescope \citep{Tyson2002SPIE} will find more of these transients.


\subsection{On the nature of GRB 111209A}
\label{SectDiscNature}
The extreme duration of the prompt emission of GRB 111209A as well as the shape of the X-ray light curve and the early optical data led to the speculation that this event might be similar to low-luminosity XRF 060218/SN 2006aj, to the ``Christmas Burst'' GRB 101225A or the relativistic tidal disruption flares (RTDFs) GRB 110328A/Swift J164449.3+573451 and Swift J2058.4+0516 (\citealt{Gendre2013ApJ}, L14). From our data and analyses, as given in G15, K18B and in this work, we find multiple large differences to these events which point to GRB 111209A not being related to these types of gamma-ray transients.

The energetics as well as the spectral parameters of the prompt emission are, excepting the extreme duration, typical for long GRBs \citep{Golenetskii2011GCN}, even the light curve shape, while strongly stretched, is similar to a typical bright, multi-peaked GRB. This is in stark contrast to the very soft and temporally simple prompt emission of XRF 060218 or XRF 100316D. Note, though, that the extreme duration implies a variability timescale that is more comparable to the aforementioned XRFs than to typical high-luminosity GRBs.
GRB 111209A is followed by an afterglow which is, while showing some complicated temporal and spectral evolution, in general very similar to typical GRB afterglows. We have derived that the optical luminosity lies close to the mean of the known afterglow luminosity distribution (K18B), and the SED shows a nearly straight and quite red spectral slope with low extinction (K18B), again fully in agreement with normal afterglows, and very different from the evolving thermal spectrum seen in the afterglow of GRB 101225A \citep{Thoene2011Nature}, or the very blue and hot thermal spectrum seen for RTDF Swift J2058.4+0516 (\citealt{Cenko2012ApJ}; and implied for GRB 110328A/Swift J164449.3+573451, which was obscured behind significant rest-frame extinction, \citealt{Levan2011Science}). The event is also in agreement with typical GRBs in the optical luminosity vs. isotropic energy release plane \citep{Kann2010ApJ}.
We detect a slow (in comparison with typical GRB-SNe), very luminous, and exceptionally blue supernova following the GRB, which agrees very well with the prototypical GRB-SN 1998bw in terms of temporal evolution, though it looks quite different from typical GRB-SNe spectrally (G15). GRB 101225A was probably also followed by a SN (which must be more luminous than the value given in \citealt{Thoene2011Nature}, in consideration of the higher spectroscopic redshift, L14), and no SN is expected in the case of RTDFs. L14 already came to the conclusion that it is unlikely that GRB 111209A originated in a RTDF, but since their data are ambiguous in terms of the detection of a supernova signature, they were not able to rule out the RTDF origin -- whereas we feel we are now able to do so in the light of our luminous SN signature as well as the ``standard'' GRB afterglow (K18B, see also G15).
The offset from the core of the host galaxy is very small (especially as measured by L14), which is in agreement with a RTDF model, but, as L14 stated, is also not in disagreement with a GRB origin (e.g., from a nuclear starburst in a very compact dwarf host galaxy).

\cite{Evans2014MNRAS} have presented a detailed discussion of the prompt emission and afterglow (in X-rays) of GRB 130925A, which we have also studied concerning its (optical)/NIR emission (K18B). They also found strong evidence that GRB 130925A is not due to a TDE/RTDF. In the GRB context, they explained the long prompt emission (see the appendix of K18B for more details) as being due to a very low circumburst medium density, this implies a very large deceleration radius $R_d$ of the jet. Shells with differing Lorentz factors therefore have more time to collide, producing internal shocks and therefore prompt gamma-ray emission, before reaching $R_d$ and the forward shock. Shells that arrive at even later times are not able to interact any more and just contribute to the external forward shock in the form of energy injections, of which we have actually detected several in the late-time afterglow of GRB 111209A (K18B). We have also found possible evidence of a very low circumburst medium density in the case of GRB 111209A (K18B, but see the arguments in \citealt{Gompertz2017ApJ}).

While this explanation therefore seems attractive also in the case of GRB 111209A, we do point out several differences to GRB 130925A. For one, the forward shock was extremely faint for the latter burst, the ``afterglow'' can be explained entirely by a dust echo \citep{Evans2014MNRAS}. While these authors state that the case is similar for GRBs 111209A and another ULGRB, GRB 121027A (see appendix of K18B), they do not find significant spectral evolution in the X-ray afterglows of the two latter GRBs, pointing to a standard external-shock origin of their afterglows. Furthermore, we have found (K18B) that the optical afterglow of GRB 111209A, while not being highly luminous, is comparable to those of many other GRBs -- and even to some with extremely luminous prompt emission, see Fig. 8 of \cite{Kann2010ApJ} for examples of GRBs with large $E_{\rm iso}$ and, relatively speaking, faint afterglows. This is also true for the X-ray afterglow \citep{Gendre2013ApJ}. The detection of bright radio emission several days after the GRB \citep{Hancock2012GCN} also indicates that the CBM density cannot be exceptionally low.

A second difference stems from the prompt emission pulses. GRB 130925A, as shown by \cite{Evans2014MNRAS}, exhibits a very large number ($\sim40$) of pulses, in general, the GRB resembles typical prompt emission, just ``lots of it''. GRB 111209A, on the other hand, shows a very slowly varying envelope. Such extremely slow variation would point to a large emission region (and thus a low circumburst-medium density, so that shells collide at large radii), as shown by \cite{Barnacka2014ApJL}, even more extreme than in the case of GRB 130925A. This very large ``Minimum Time Scale'' (MTS) would also point to a very low Lorentz factor \citep{Sonbas2015ApJL}, but this is incompatible with opacity arguments (recall that the peak energy of the prompt emission is $>500$ keV). On the other hand, the MTS of GRB 111209A likely lies significantly above all the values plotted in \cite{Sonbas2015ApJL}, so it is unclear how the relationship of these authors would apply.

\cite{Margutti2015ApJ} posited a direct physical connection between extremely long prompt emission duration and high circumburst X-ray column densities as well as very soft late X-ray afterglows which indicate reprocessing of X-ray emission from the forward shock. They only studied GRBs at $z<0.5$ and found four examples, the XRFs 060218 and 100316D as well as the two GRBs 090417B and 130925A (both \emph{bona fide} dark GRBs). Due to the redshift cutoff, GRB 111209A was not included. This burst shares some of the aspects of their sample (obviously, the ultra-long duration, as well as the very long variability timescale), but is markedly different in others. Using the tools of the XRT online repository \citep{Evans2007AA,Evans2009MNRAS}, we find that the X-ray afterglow, using only data after 100 ks, has a somewhat softer spectrum ($\Gamma=2.50^{+0.20}_{-0.18}$) than expected for a forward-shock-driven afterglow, but is not in the same region as the Margutti sample (all $\Gamma>3$); also, as mentioned above, there is no spectral evolution detected. The equivalent hydrogen column density as measured from X-rays, $N_H=2.5^{+1.1}_{-1.0}\times10^{21}$ cm$^{-2}$, is typical for GRB afterglows, and as we show in K18B, the line-of-sight extinction in the optical is low. GRB 111209A does lie in the upper right hand quadrant of Figure 2 of \cite{Margutti2015ApJ} as do the other very long GRBs they have studied, but it lies significantly below the rest of the sample. Furthermore, the prompt emission of GRB 111209A shows a much higher peak energy than any of the cases in the sample of \cite{Margutti2015ApJ}. Therefore, if there is indeed a physical link between an extremely long prompt-emission duration and the pre-explosion mass-loss history of the progenitor, it seems less likely that this link also applies to the case of GRB 111209A, with the extreme duration more likely to be intrinsic than a result of a complex mass-loss environment \citep{Margutti2015ApJ}.

Thanks to the classification of the spectrum of SN 2011kl as being driven by the energy release from a magnetar (G15), it would seem we can narrow down the possibilities, and the extreme duration of the prompt emission could be directly linked to the existence of a magnetar central engine. The exact connection remains unclear, though. Magnetars as central engines of GRBs have been invoked multiple times. XRF 060218 has been invoked as an example of a magnetar-powered GRB \citep{Mazzali2006Nature}, but its prompt emission, while very long, differs strongly from that of GRB 111209A, and the associated SN 2006aj shows no signs of being anything but a Type Ic-BL SN powered by radioactive decay \citep[e.g.,][]{Pian2006Nature}. Magnetars have also been used to explain a rarely seen phenomenon of a long-lasting X-ray plateau followed by a very steep decay at late times \citep{Zhang2001ApJ}, with the optical afterglow behaving completely differently, prime examples being GRBs 070110 \citep{Troja2007ApJ} and 130831A \citep{DePasquale2016MNRAS}. These GRBs were otherwise completely unremarkable, though; GRB 130831A was also followed by a ``garden-variety'' Type Ic-BL SN, SN 2013fu (\citealt{Cano2014AA,Klose2018AA}). Therefore, even though this paper, together with G15 and K18B, shows strong evidence for a GRB central engine not involving a rapidly spinning black hole, the exact connection between the ultra-long duration and the spectrally deviant, luminous SN remains a topic for further research.

\subsection{The literature on GRB 111209A in light of our complete data set}
\label{SectDiscLit}

GRB 111209A and long-lasting gamma-ray transients in general have been much-discussed in the literature in recent years. Most of the studies pertaining to GRB 111209A were based on significantly incomplete data sets.

\subsubsection{Models for gamma-ray transients of extreme duration}

\cite{Quataert2012MNRAS} and \cite{Woosley2012ApJ} studied the possibility of explaining long-duration $\gamma$-ray transients, such as the ones GRB 111209A has been postulated to resemble (Sect. \ref{SectDiscNature}), through the long-term accretion of the outer layers of massive stars, generally ones that have \emph{not} experienced envelope-stripping. Such transients are unable to explain GRB 111209A. Most models lead to low-luminosity emission which lasts dozens, if not hundreds of days. Hereby, the emitted luminosity would lie one to two orders of magnitude below that of GRB 111209A. While a BSG progenitor would lead to an emission time in agreement with GRB 111209A \citep[$10^4-10^5$ s,][]{Woosley2012ApJ}, it is expected to be even less luminous than a red supergiant progenitor. Furthermore, these ``Type 3 Collapsars'', as \cite{Woosley2012ApJ} label them, would also very likely not be accompanied by supernova emission (as such a supernova would detach exactly the outer layers needed for accretion), in contrast to the bright SN we have discovered.

\cite{Quataert2012MNRAS} also contemplate if a millisecond magnetar central engine, which in general has been posited to be a valid central engine model for GRBs \citep[e.g.,][and references therein]{Metzger2011MNRAS}, is able to power a long-duration low-luminosity $\gamma$-ray transient, and find that the magnetic field strength must lie under that of typical magnetars to enable a much slower spin-down and therefore a longer emission period. Again, such a transient would be much longer and fainter than GRB 111209A, but there is no argument against scaling up the magnetic field to a value between the ``classic'' millisecond magnetar central engine which emits most of its energy on the time scale of typical long GRBs, and the scenario which \cite{Quataert2012MNRAS} propose. Such a scenario is discussed in the light of the spectrum of SN 2011kl presented in G15, though caveats in terms of variability and energetics remain (Sect. \ref{SectDiscNature}, see also, e.g., \citealt{Gompertz2017ApJ}).

\cite{Janiuk2013AA,Janiuk2013conf} propose a scenario in which the exploding progenitor is in a tight orbit with a second stellar-mass black hole. Shortly after exploding as a supernova, the two black holes merge, and additional accretion powers a second episode of ultrarelativistic jet launching, leading to a very long-duration GRB. While such a model may explain ``double bursts'' such as GRB 110709B \citep{Zhang2012ApJ} and GRB 121217A \citep{Siegel2013GCNR, Elliott2013AA}, it is unlikely to explain the sustained and extremely extended, but low-peak-flux emission of GRB 111209A.

\cite{Nathanail2015MNRAS} present a model in which ultra-long GRBs are explained by delayed accretion on to a black hole engine, the delay being achieved by a lower-than-usual magnetic field strength. Such a model would be in contrast to the magnetar scenario (G15), but \cite{Nathanail2015MNRAS} concede that is possible for a magnetar to work within their model. 

\cite{Gilkis2016ApJ} study core-collapse SNe in general within the framework of the jittering-jets model, and propose that for progenitors with very high pre-collapse angular momentum (generally seen as a prerequisite for GRBs), strong collimated polar outflows are created, as well as a slow equatorial outflow which partially forms a massive and extended accretion disk. Such a disk, they propose, could continue the jet-emission process, thereby powering the SN to SLSN luminosities. Furthermore, if the accretion time is measured in hours to days, it would yield a natural explanation for ultra-long duration GRBs, and possibly also for the high luminosity of an associated SN, as well as for the fast-rise gap transients \cite{Arcavi2015ApJ} have studied. It remains to be seen whether such a model would yield the specific spectral shape of SN 2011kl, though (G15). We furthermore note that the model of \cite{Gilkis2016ApJ} predicts definite asymmetries in the explosion, {but polarimetry studies of SLSNe so far have yielded null results \citep{Leloudas2015ApJL,Brown2016ApJ,Cikota2018MNRAS,Maund2019MNRAS}, with some exceptions \citep{Inserra2016ApJ,Leloudas2017ApJ,Bose2017ApJ}}.

\cite{Perets2016ApJ} discuss what they call ``Micro Tidal Disruption Events'' ($\mu$TDEs), in which planetary mass objects or solar-mass stars are tidally disrupted by \emph{stellar mass} black holes. They find that in the case of solar-type stars, where the mass ratio is $\approx0.1$ (or even higher), jet production may set in, and a long-lasting gamma-ray flare will be produced, whose duration and energetics agree well with ultra-long GRBs. While $\mu$TDEs in general would not produce any SN emission, they also envision a special scenario in which a massive star explodes as a SN, producing the compact object, whose natal kick brings it close to a wide companion, thereby producing the $\mu$TDE just hours or days after the SN detonation. They mention GRB 111209A/SN 2011kl as a possible candidate for such a scenario. Still, there is no obvious way to explain why SN 2011kl has the measured properties, in terms of luminosity and spectral shape. Furthermore, it is unclear if such a scenario can produce a classical GRB jet which then causes a quite standard afterglow as in the case of SN 2011kl. The agreement of SN 2011kl with the SN 1998bw template indicates that GRB 111209A and the SN are likely contemporaneous, but a delay of a few hours (in this case, the SN would precede the GRB!) is not in disagreement with the data. \cite{Perets2016ApJ} find that black holes make better compact objects for the creation of $\mu$TDEs, but a NS, and possibly even a magnetar, is not ruled out. One may even envision that the secondary accretion event could spin up the NS further, increasing the energy reservoir needed to power the SN to its high luminosity. Therefore, while the scenario seems very fine-tuned, it remains an interesting candidate which should be further explored.

\cite{Gao2016ApJL} present an alternate model in which GRB 111209A is powered by fall-back accretion on to a black hole central engine, which would explain the ultra-long duration. Furthermore, the fall-back accretion disc should then power the high luminosity (and, possibly, different spectral shape compared to the usual Type-IC BL associated with GRBs) of SN 2011kl via the Blandford-Payne mechanism \citep{Blandford1982MNRAS}. It remains unclear, though, whether such a mechanism can indeed reproduce the spectral peculiarities of SN 2011kl, and whether a black hole fall-back process can extend long enough, after powering the ultra-long GRB, to also yield the power to produce the high SN luminosity \citep{Metzger2015MNRAS}.

{\cite{Beniamini2017MNRAS} study the ability of magnetars, assuming different emission models, in powering GRBs. They find that $T_{90}$ should typically be $\approx{100}$ s, in strong contrast to the existence of ULGRBs. They do point out, though, that the prompt emission time may be enhanced if the jet entrains a significant baryon loading when passing through the star. Additionally, further fallback accretion may increase the energy output of the magnetar and produce a longer emission timescale as well.}

{The case of fallback accretion on a proto-magnetar is further studied by \cite{Metzger2018ApJ}, who indeed find that under certain circumstances, such fallback can hold the magnetization at a critical level to enable prompt emission times in the range of thousands of seconds, while at the same time still being able to power the accompanying SN to luminosities exceeding those achievable by $^{56}$Ni decay alone (such a proto-magnetar would also produce typical amounts of $^{56}$Ni right after core-collapse). This model therefore represents one of the best solutions for GRB 111209A/SN 2011kl presented so far.}

{\cite{Liu2018ApJ} use prompt emission data to constrain the possible progenitor stars of GRBs of different duration within the framework of a Black-Hole hyper-accretion model. They find that ULGRBs such as GRB 111209A can only have progenitors of high mass and low metallicity, up to metal-free Pop III stars. While this model is similar to the standard model and therefore an accompanying SN is expected, it does not explain the specific properties of SN 2011kl, which point to a magnetar origin.}

{\cite{Perna2018ApJ} use multi-code numerical modelling to evolve a low-metallicity, massive star to a BSG end phase, and then produce a long-lasting, ultrarelativistic jet which is able to drill through the extended envelope and produce an ULGRB of $\approx10$ ks duration if seen near axis. While this modelling shows that BSG progenitors are able to produce ULGRBs, it does not encompass the creation of the SN or explain its high luminosity and spectral properties. Note also this model uses a black hole central engine.}

{\cite{Aguilera-Dena2018ApJ} present evolutionary models of potential Type Ic progenitor stars that reach large C/O-core masses via enhanced mixing. Depending on the core mass, they find a continuum spanning from Type Ic SLSNe over magnetar-powered GRB-SNe, black-hole powered GRB-SNe up to PPI-SNe. The potential magnetar-powered GRB-SNe are found near the NS/BH boundary, implying very massive, rapidly rotating neutron stars (see K18B for more discussion about the cumulative energetics of the event). \cite{Aguilera-Dena2018ApJ} mention this part of the paramter space as a potential progenitor for GRB 111209A.}

\subsubsection{GRB111209A: then and now}

\cite{Gendre2013ApJ} present a detailed study of the GRB 111209A prompt emission, mostly in $\gamma-$rays and X-rays. They rule out an origin of GRB 111209A in a supernova shock breakout (lack of a strong thermal component, energetics), a magnetar origin (discrepancy between energy release and prompt-emission peak energy, light curve behaviour in X-rays), and a RTDF origin (light curve behaviour in X-rays). Concerning the latter, we note that \cite{Gendre2013ApJ} also cite the lack of a host galaxy as a counterargument against a RTDF origin, whereas both L14 and K18B present host-galaxy detections, and L14 argue that the host is still massive enough to contain a central supermassive black hole that can create a RTDF. Finally, \cite{Gendre2013ApJ} come to the conclusion that the only model supported by their data is that of the core collapse of a single low-metallicity supergiant star, a BSG. They base this decision on the apparent lack of any supernova detection for GRB 111209A, as, at the time of writing of their manuscript, only a preliminary analysis of the data set presented in L14 had been made public. The final analysis of L14 revealed indications of an accompanying SN, and our own observations (G15, K18B) show that not only does the SN signature exist, it is highly luminous, in contrast to the expectations of the model of \cite{Woosley2012ApJ} and therefore \cite{Gendre2013ApJ}, as already stated above. Additional arguments against a low-metallicity BSG progenitor are made by L14 (host-galaxy metallicity, but see \citealt{Kruhler2015AA}) and \citet[][detection of dust along the GRB line-of-sight]{Stratta2013ApJ}, the latter argument is also confirmed in K18B. \cite{Gao2016ApJL}, while favouring a Black-Hole-powered model, also rule out an extended star as a progenitor.

\cite{Kashiyama2013ApJ} have developed models for an optical transient phenomenon they label Cocoon Fireball Photospheric Emission (CFPE), which can create luminous, long-lasting transients peaking in the optical regime from massive BSG progenitors (e. g., Luminous Blue Variables or even Population III stars). Due to the link to a possible BSG progenitor for GRB 111209A that \cite{Gendre2013ApJ} had posited, \cite{Kashiyama2013ApJ} have also modelled what such a transient would look like at the redshift of GRB 111209A (their Figure 6). It would peak around 26th - 25th magnitude (depending on the zero-age main sequence mass of the progenitor), with a peak time of $\approx2-2.5\times10^{7}$ s (several hundred days) and a very blue spectrum (rising toward the ultraviolet, so the opposite of an afterglow or typical SN spectrum). Together with the late observations of L14, we have only two epochs spanning this late time. No significant variability is detected in $g^\prime r^\prime$ between 200 and 280 days. We note, though, that the transient \cite{Kashiyama2013ApJ} find in their models has a roughly symmetrical magnitude evolution in log space, and may therefore peak in between the two epochs, the lack of variability being a chance effect of the temporal spacing. Therefore, we are unable to significantly exclude the existence of such a transient. The additional component we find at $\approx$ tens of days after the GRB is much faster and much more luminous than the CFPEs \cite{Kashiyama2013ApJ} find in their model, and therefore very unlikely to be due to this phenomenon. Furthermore, as we have argued beforehand, it is unlikely that the progenitor is a BSG, and therefore the model of \cite{Kashiyama2013ApJ} would not be applicable anyway. They also derive models for CFPEs of exploding Wolf-Rayet stars, but these are several magnitudes fainter than the BSG CFPEs and would be completely undetectable in our data, being far less luminous than even the faint host galaxy of GRB 111209A.

\cite{Nakauchi2013ApJ} expand upon the work of \cite{Kashiyama2013ApJ}, fitting new CFPE models to the data presented by L14. They find that using a certain set of initial parameters, they are able to reproduce the red rebrightening seen especially in the $J$ band data of L14, which they claim to be about an order of magnitude more luminous than typical GRB-SNe, and therefore SLSN-like. While we find the observer-frame $J$-band SN to be several times more luminous than SN 1998bw in the same band, this is still below the regime of SLSNe. Comparing our data with the model light curves of \cite{Nakauchi2013ApJ} shows that their model significantly exceeds our data in luminosity, and additionally there is no evidence for the very steep decay they find. Therefore, if the rebrightening following GRB 111209A is to be explained by such a modified CFPE, the initial parameters must be very different from those chosen by \cite{Nakauchi2013ApJ}. Note that they do agree with \cite{Kashiyama2013ApJ} in that CFPEs of Wolf-Rayet stars would be too faint to detect, such an effect is clearly ruled out by our data.

\cite{Ioka2016ApJ} apply three different models to the GROND data of GRB 111209A/SN 2011kl: A blue supergiant model, a magnetar model and a model involving a tidal disruption flare caused by a relatively low-mass supermassive BH destroying a white dwarf. They claim that all three models can fit the data, but favor especially WD-TDF model, and find problems with the parameters of the magnetar model. In their paper, they only employ the GROND data as presented in G15, thereby ignoring the rebrightening episode and the complex early afterglow, as well as the multiple smaller rebrightenings in the late afterglow. Furthermore, they do not show residuals of their fits or give $\chi^2$ values. Visual inspection of their fits reveals strong offsets between data and fit curves in some colours, and it is unclear whether their models can match the spectrally determined colours of SN 2011kl. We therefore conclude that as it stands, their modelling does not present a strong argument against the magnetar model for SN 2011kl. We also note that other authors, e.g., \cite{Yu2017ApJ}, have derived magnetar-model parameters which are unexceptional and in agreement with those of, e.g., SLSNe (see Sect. \ref{Power}). {\cite{Wang2017ApJ3} have also used the bolometric light curve decomposition of \cite{Ioka2016ApJ} and find it can be fit with a $^{56}$Ni+magnetar+cooling envelope model, without the need for a BSG or WD-TDE model.}


\section{Conclusions}
\label{SectConc}


In a series of papers, we have studied the ultra-long GRB 111209A. In G15, we presented the discovery of the highly luminous SN 2011kl accompanying GRB 111209A, a remarkable event that is spectrally more similar to SLSNe. In K18B, we studied the entire optical/NIR afterglow of GRB 111209A, finding a complex evolution but, in contrast to SN 2011kl, no features that set it apart from the known sample of GRB afterglows.

In this work, we studied SN 2011kl within the context of large SE-SN, GRB-SN and SLSN samples. We derive a bolometric light curve including NIR corrections, confirming and expanding upon the results of G15 that this SN would be powered by $^{56}$Ni, it would be a significant outlier compared to all known (non-SLSN) SE-SNe. At the same time, in terms of luminosity and rapidity of light-curve evolution, SN 2011kl is also an outlier compared to the SLSN sample which it resembles spectroscopically (G15, \citealt{Mazzali2016MNRAS}), being less luminous and faster, and inhabiting the ``luminosity gap'' between GRB-SNe and SLSNe, in which only a few more sources have been found so far \citep[e.g.,][]{Arcavi2015ApJ}{, some of them being spectrally confirmed SLSNe \citep{Decia2017ApJ,Lunnan2018ApJ,Quimby2018ApJ}}. All in all, SN 2011kl is a true ``hybrid'' and represents a transition object between the rapidly expanding SE-SNe accompanying GRBs and the highly luminous SLSNe.

Even in the light of our studies, a complete explanation for GRB 111209A/SN 2011kl is still lacking. Is a magnetar capable of powering the entire event \citep[][K18B]{Gompertz2017ApJ}? Is this truly necessary or are certain elements, such as the afterglow, powered by ``conventional'' mechanisms known from standard GRBs (e.g., deceleration by the circumburst medium and synchrotron radiation from a forward shock in the case of the afterglow)? Is the ultra-long prompt emission phase related to accretion of similar duration or is it produced by interaction with the environment \citep{Evans2014MNRAS}?

So far, very few ultra-long duration GRBs are known, and this is the first one exhibiting an accompanying SN, which, at first glance, links it to ``normal'' long GRBs. A more detailed analysis has revealed several striking differences, though, indicating that we as yet have not discovered the entire ``bestiary'' of gamma-ray transients in the Universe. Any further similar events in the future need to be followed up with all possible effort, but their low gamma-ray peak fluxes and long-scale variability may bias these events against detection, especially at higher redshifts.

Lately, evidence has been growing that there is a continuum of SE-SNe reaching from Types IIb, Ib, Ic over Type Ic-BL and GRB-SNe \citep[e.g.,][]{Mazzali2008Sci}, possibly due to common jet physics {(e.g., \citealt{Piran2017Sci,Sobacchi2017MNRAS1,Petropoulou2017MNRAS,Soker2017ApJ})}
all the way to SLSNe-I (possibly due to envelope properties, \citealt{Sobacchi2017MNRAS2}, magnetar-axes alignment, \citealt{Margalit2017MNRAS,Margalit2018MNRAS}, magnetar field strength, \citealt{Yu2017ApJ}{, or C/O core mass at explosion time, \citealt{Aguilera-Dena2018ApJ}}). Similarities have been found in the spectral properties both in the photospheric \citep{Liu2017ApJ,Blanchard2019ApJ} and the nebular \citep[e.g.,][]{Milisavljevic2013ApJ,Nicholl2016ApJ2,Jerkstrand2017ApJ} phases, {spectroscopically classified SLSNe have been discovered to populate the luminosity ``gap'' \citep{Decia2017ApJ,Lunnan2018ApJ,Quimby2018ApJ,Angus2019MNRAS}}, X-ray observations reveal similar environments \citep{Margutti2017ApJX}, and there are indications from simulations \citep{Suzuki2017MNRAS} as well. {All in all, there are more and more indications for the existence of a connection between GRBs and SLSNe \citep{Margalit2017MNRAS}. While a connection between SLSNe and highly-stripped progenitors was quickly established \citep{Pastorello2010ApJ}, a connection to GRBs was hardly obvious and has been forthcoming more slowly. GRB 111209A/SN 2011kl may ultimately turn out to be the first ``missing link'' between classical GRBs and SLSNe to be discovered; in a case where it was not, at the time, suspected that such a missing link was even needed. The final step would now be to discover a high signal-to-noise, spectroscopically indubitable SLSN (even if it doesn't make the ``luminosity cut'') which is clearly associated with a GRB, be it of ultra-long duration or not.}


\acknowledgements
DAK wishes to dedicate these works to his father, R.I.P. 20. 08. 2015. You are sorely missed by so many. DAK acknowledges Zach Cano, Massimiliano De Pasquale, Daniele Malesani, Antonio de Ugarte Postigo, Christina C. Th\"one, Bing Zhang, Thomas Kampf, Cristiano Guidorzi, Raffaella Margutti, and Ting-Wan Chen for interesting discussions and helpful comments. {DAK also thanks A. Sagues Carracedo for further information on the bolometric light curve of SN 2017iuk.}
DAK acknowledges financial support by the DFG Cluster of Excellence ``Origin and Structure of the Universe,'' from MPE, from TLS, from the Spanish research project AYA 2014-58381-P, and from Juan de la Cierva Incorporaci\'on fellowship IJCI-2015-26153.
We are indebted to Joe Lyman and Vicki Toy for supplying the bolometric light curves of GRB 120422A/SN 2012bz and GRB 130702A/SN 2013dx, respectively.
SK, DAK, ARossi, and ANG acknowledge support by DFG grants Kl 766/16-1 and Kl 766/16-3, SSchmidl also acknowledges the latter.
ARossi acknowledges support from the Jenaer Graduiertenakademie and by the project PRIN-INAF 2012 ``The role of dust in galaxy evolution''.
TK acknowledges support by the DFG Cluster of Excellence Origin and Structure of the Universe, and by the European Commission under the Marie Curie Intra-European Fellowship Programme.
RF acknowledges support from European Regional Development Fund-Project ``Engineering applications of microworld physics'' (No. CZ.02.1.01/0.0/0.0/16\_019/0000766).
DARK is funded by the DNRF.
FOE acknowledges funding of his Ph.D. through the DAAD, and support from FONDECYT through postdoctoral grant 3140326.
SSchulze acknowledges support from CONICYT-Chile FONDECYT 3140534, Basal-CATA PFB-06/2007, and Project IC120009 ''Millennium Institute of Astrophysics (MAS)'' of Iniciativa Cient\'ifica Milenio del Ministerio de Econom\'ia, Fomento y Turismo.
SK, SSchmidl, and ANG acknowledge support by the Th\"uringer Ministerium f\"ur Bildung, Wissenschaft und Kultur under FKZ 12010-514.
MN and PS acknowledge support by DFG grant SA 2001/2-1.
ANG, DAK, ARossi and AU are grateful for travel funding support through MPE.
Part of the funding for GROND (both hardware as well as personnel) was generously granted from the Leibniz-Prize to Prof. G. Hasinger (DFG grant HA 1850/28-1).
This work made use of data supplied by the UK Swift Science Data Centre at the University of Leicester.

\begin{appendix}
\section{Data sources for Fig. \ref{SLSNeBol}}
\label{Fig4data}
GRB-SNe: \cite{Olivares2012AA,Olivares2015AA,Schulze2014AA,Toy2015ApJ,Ashall2017Nature,Cano2017AA,Corsi2017ApJ,deUgartePostigo2017GCN,Izzo2019Nature}.

Gap transients and others: \cite{Arcavi2015ApJ,Roy2016AA,Vallely2017MNRAS,Whitesides2017ApJ,Benetti2018MNRAS,Chen2018ApJL}, Kann et al., in prep.

SLSNe: \cite{Quimby2011Nature,Chomiuk2011ApJ,Rest2011ApJ,Leloudas2012AA,Cooke2012Nature,Chatzopoulos2013ApJ,Kostrzewa-Rutkowska2013ApJ,Howell2013ApJ,Nicholl2013Natur,Nicholl2014MNRAS,Nicholl2015ApJL,Nicholl2016ApJ,Nicholl2017ApJ,Lunnan2013ApJ,Lunnan2016ApJ,Lunnan2018ApJ,Lunnan2018NatAs,Inserra2013ApJ,Inserra2018MNRAS,Inserra2017MNRAS,Benetti2014MNRAS,Vreeswijk2014ApJ,Vreeswijk2017ApJ,McCrum2013MNRAS,McCrum2015MNRAS,Papadopoulos2015MNRAS,Yan2015ApJ,Yan2017ApJ2,Smith2016ApJ,Chen2017AA,Decia2017ApJ,Blanchard2018ApJ,Blanchard2019ApJ,Anderson2018AA,Taddia2019AA2,Angus2019MNRAS}.

ASASSN-15lh: \cite{Dong2015Sci}.

\section{Erratum to K18B}
In K18B, we measured the offset between GRB 111209A/SN 2011kl and the centre of its host galaxy. Hereby, the correction for the cosine of the rectascension was not implemented. With this additional term, we find an offset in RA of 0\farcs271, and therefore a total offset of $0\farcs33\pm0\farcs18$, which translates to a projected offset of $2.26\pm1.25$ kpc. This is in better agreement with the result of L14, and otherwise does not change our conclusions.
\end{appendix}

\bibliographystyle{aa}

\end{document}